\documentclass[aps,prl,twocolumn,floatfix,superscriptaddress]{revtex4-2}
\usepackage[utf8]{inputenc}
\usepackage[T1]{fontenc}
\usepackage[english]{babel}
\usepackage{mathtools}
\usepackage{newtxtext}
\usepackage{amsmath}
\usepackage{amsfonts}

\usepackage{booktabs}

\renewcommand\vec{\mathbf}

\usepackage[dvipsnames,svgnames,x11names,hyperref]{xcolor}
\usepackage{hyperref}
\hypersetup{
    colorlinks=true, linkcolor=NavyBlue, urlcolor=NavyBlue, citecolor=NavyBlue
}

\usepackage{physics,braket,siunitx,slashed}
\usepackage[capitalise]{cleveref}

\usepackage{tabularx}

\newcommand{\bk}{\mathbf{k}}
\newcommand{\br}{\mathbf{r}}
\newcommand{\bG}{\mathbf{G}}
\newcommand{\bQ}{\mathbf{Q}}
\newcommand{\bq}{\mathbf{q}}
\newcommand{\bkbar}{\bar{\bk}}
\newcommand{\bqbar}{\bar{\bq}}

\begin{document}

\title{Retarded Interaction Between Opposite Chiral Edges in Anomalous Hall Crystals}

\author{Wangqian Miao}
\affiliation{Department of Physics, The Pennsylvania State University, University Park, Pennsylvania 16802, USA}
\affiliation{Center for Theory of Emergent Quantum Matter,
The Pennsylvania State University, University Park, Pennsylvania 16802, USA}

\author{Mu-Yang Chen}
\affiliation{Department of Physics, The Pennsylvania State University, University Park, Pennsylvania 16802, USA}

\author{Binghai Yan}
\affiliation{Department of Physics, The Pennsylvania State University, University Park, Pennsylvania 16802, USA}
\affiliation{Center for Theory of Emergent Quantum Matter,
The Pennsylvania State University, University Park, Pennsylvania 16802, USA}

\author{Chunli Huang}
\affiliation{Department of Physics and Astronomy, University of Kentucky, Lexington, Kentucky 40506-0055, USA}

\date{\today} 

\begin{abstract} An anomalous Hall crystal combines spontaneous electronic crystallization with a Chern insulating gap, supporting both chiral edge modes and low-energy electronic phonons. We show that this coexistence produces a distinct dynamical effect from ordinary Chern insulators: transverse bulk phonons can mediate a retarded interaction between counterpropagating chiral edge modes on opposite sides of the sample, realizing a Luttinger-liquid variant with delayed inter-edge coupling. Using microscopic time-dependent Hartree--Fock calculations for rhombohedral pentalayer graphene, we find that lowering the carrier density softens the long-wavelength transverse phonon mode. Near this instability regime, the resulting boundary-projected phonon continuum inevitably overlaps with the edge dispersion, thereby enabling their coupling. A smoking-gun probe is a nonlocal measurement: a drive applied to one edge can induce a response on the other edge, delayed by the transverse phonon time of flight across the sample.
\end{abstract}

\maketitle

\textcolor{NavyBlue}{\textit{Introduction.}} Electronic crystals \cite{Wigner_original, Needs_HFWC, Bernu_MIT_2008, sandeep25chiral, bfield_wc_kim} that intertwine topology with broken translational symmetry with no magnetic field, dubbed anomalous Hall crystals \cite{tesanovic_hall_1989_prb,dong_anomalous_2024, dong_stability_2024, zeng_sublattice_2024, soejima_anomalous_2024, 
tan_parent_2024, dong_theory_2024, zhou24moireless, patri24extended, 
zhou_new_2025, guo2025ahcblg, zeng25berryslide,tomohiro2025jellium,tan_ideal_2025,miao2026ahc,uchida2026nonabilien,Desrohcers26lamdaN}, 
 are among the latest additions to the zoo of correlated states reported in rhombohedral graphene \cite{lu_fractional_2024,lu_extended_2025,aronson_displacement_2025}.  Like Chern insulators, AHCs host chiral edge modes at the boundary that give rise to the quantum anomalous Hall effect. At the same time, like Wigner crystals, they have low-lying collective phonon excitations in the bulk arising from continuous translational symmetry breaking \cite{2D_BWC_Goldoni_1996, excitations_in_GWC,
 dong_phonons_2025, ambuj2025phonon, mark2026phonon}. 
 Recent transport experiments in rhombohedral graphene with an hBN moiré superlattice found that, when electrons are layer-polarized to the layer furthest from the moiré potential, the quantum anomalous Hall effect appears and, surprisingly, it persists over an extended density range at the lowest accessible temperatures \cite{lu_extended_2025}. This behavior can be explained with a candidate AHC stabilized by a weak moiré potential, although its precise nature remains to be established \cite{kwan_moire_2025,Huo_does_2025, Qiao2026-iy, Luca_relaxation}. As the density is reduced, the system appears to evolve toward a conventional Wigner crystal with large longitudinal resistance \cite{ lu_extended_2025,han2026WC,jacson_26_gaplessqah}. Understanding the connection between these phases, their relation to nearby fractional quantum anomalous Hall states, and their possible link to chiral superconductivity is one of the most challenging problems in 2D materials.

\begin{figure}
\includegraphics[width=0.95\columnwidth]{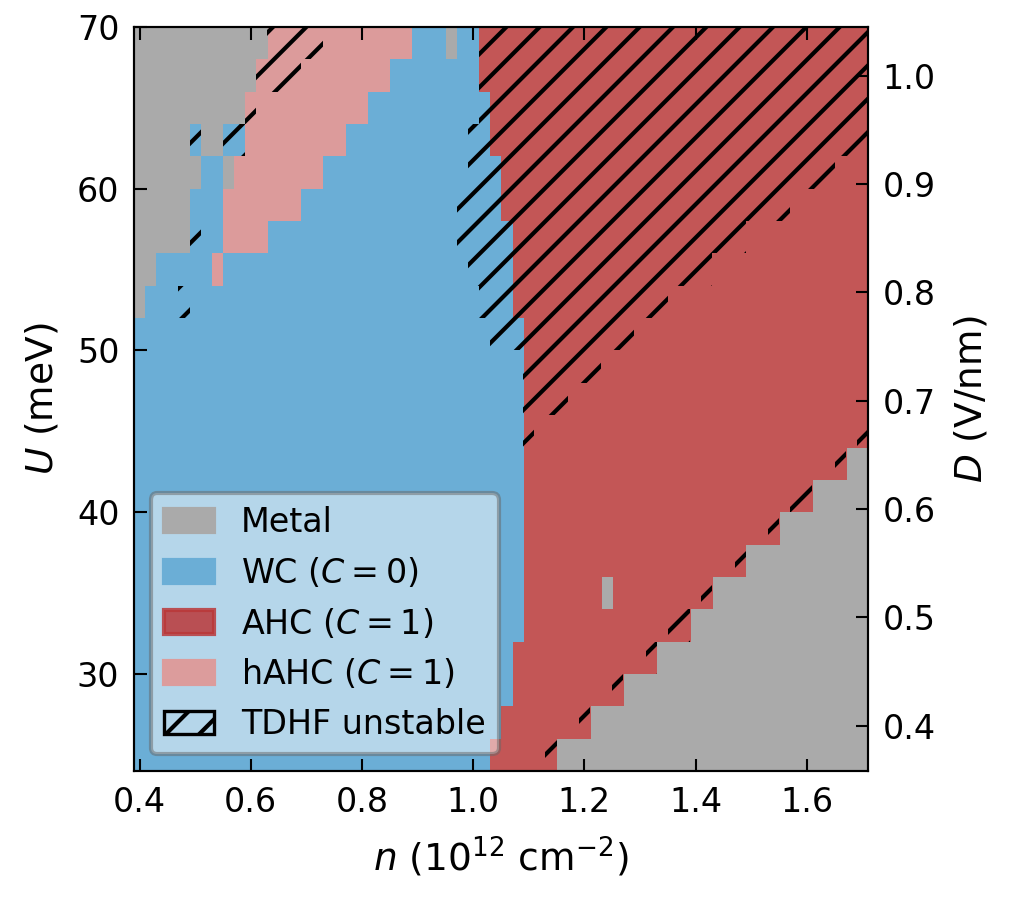}
\caption{ Hartree-Fock phase diagram of rhombohedral pentalayer graphene (R5G) as a function of carrier density ($n$) and displacement field ($D$ or $U$). The translation-symmetry-broken phases include the Wigner crystal (WC), halo anomalous Hall crystal (hAHC), and anomalous Hall crystal (AHC). All crystalline and metallic phases are fully polarized into a single spin-valley flavor, thereby breaking time-reversal symmetry. Phase stability is evaluated using the time-dependent Hartree-Fock (TDHF) approximation, with unstable regimes denoted by dashed lines. As discussed in the main text, the softest instability likely corresponds to a stripe phase.}
\label{fig1}
\end{figure}

\begin{figure*}
\includegraphics[width=1.98\columnwidth]{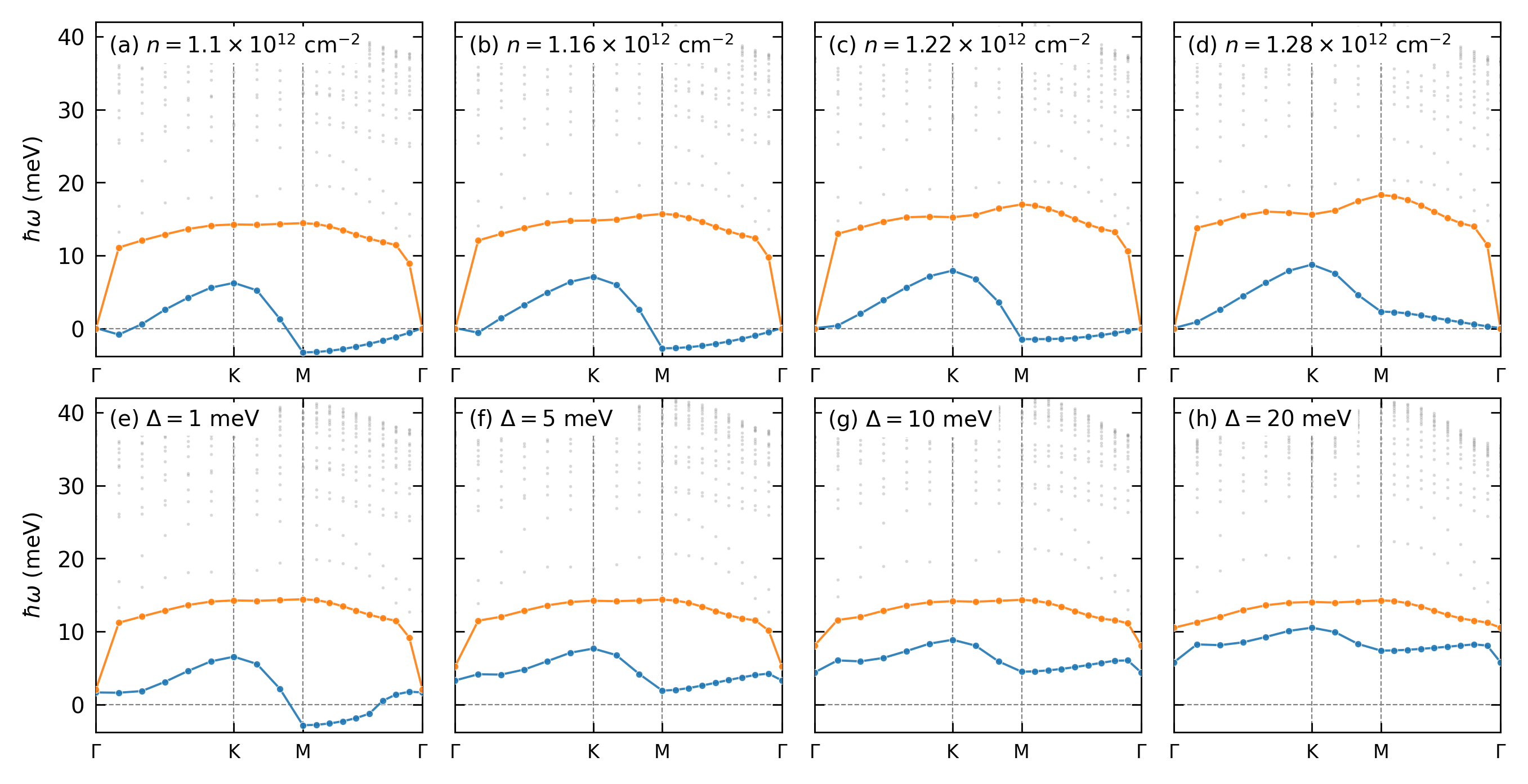}
\caption{
TDHF collective-mode spectrum as a function of carrier density at fixed displacement field. The orange branch denotes the plasmon mode, which has the expected $\sqrt{q}$ dispersion near the $\Gamma$ point. The blue branch denotes the transverse phonon mode. Negative values indicate modes with purely imaginary frequency, signaling an instability of the Hartree--Fock state. The instability is strongest near the $M$ point, where the imaginary frequency is largest. As the density is lowered toward the continuous transition, the transverse phonon mode softens near the $\Gamma$ point, producing the low-energy bulk continuum responsible for bulk--edge hybridization. We fix $U=50$ meV in these calculations. (a)-(d) The pinning potential is fixed to be $\Delta=0$ meV and $n$ is taken as a variable. (e)-(f) $n$ is fixed to be $n=1.1\times10^{12}$ cm$^{-2}$ and the pinning potential $\Delta$ is set to be a variable. All calculations are done on a $18 \times 18$ $k$-mesh.
}
\label{fig2}
\end{figure*}

Despite intense recent interest, a clear experimental smoking-gun signature that distinguishes AHCs from conventional Chern insulators remains lacking. In this Letter, we identify a universal signature of AHCs in the vicinity of a continuous phase transition. Using microscopic time-dependent Hartree–Fock calculations \cite{thouless1960stability,thouless1962time,eslam20softmode,kwan_mean-field_2025,ambuj2025phonon,dong_phonons_2025,shao25elecmag,mark2026phonon}, we show that the gapless transverse phonon mode in the bulk necessarily couples to the chiral edge mode at the boundary as the system approaches the continuous phase transition point. This coupling induces hybridization between edge excitations and long-wavelength phonons, thereby allowing phonons to mediate retarded interactions between spatially separated edges on opposite sides of the sample, even though the edge-state wavefunctions have no spatial overlap. This coupling is absent in existing Chern insulators \cite{chang2023colloquium}, (valley-polarized) moiré superlattices and magnetic topological insulators, because their low-lying collective modes are symmetry-decoupled from the charge excitation at the boundary. This distinction represents a unique feature of AHCs and provides a concrete, experimentally accessible signature that can be systematically probed.

\textcolor{NavyBlue}{\textit{Anomalous Hall crystal in rhombohedral graphene.}}
We study one of the AHC candidates, rhombohedral graphene multilayers using the standard $k \cdot p$ continuum Hamiltonian
\cite{herzog-arbeitman_moire_2024,miao2026ahc}: 
\begin{equation}
    H = \sum_\bk h(\bk)+\frac{1}{2}\sum_{\bq} v_{\ell\ell'}(\bq):\rho_{\bq\ell}\rho_{-\bq\ell'}:+V(\mathbf{r}),
\end{equation}
where $h(\bk)$ is the single particle Hamiltonian for rhombohedral graphene, $\rho_{\bq\ell}$ is the density operator for layer $\ell$ and $v_{\ell \ell'}(\bq)$ is the layer resolved screened Coulomb interaction, $V(\mathbf{r})$ is an external pinning potential which can originate from the hBN alignment \cite{herzog-arbeitman_moire_2024,dong_anomalous_2024, dong_theory_2024,  zhou24moireless, kwan_moire_2025} or an external patterned gate \cite{Ghorashi_topological_2023_prl, zeng_sublattice_2024, Miao_artificial_2025_prb,Shi_fractional_2025}. Details of the
model and numerical setup are provided in the Supplemental Material (SM) \cite{Supp}. We set the dielectric constant $\varepsilon=5$ in our calculation throughout the paper. All four
spin-valley flavors are retained, so that time-reversal symmetry is preserved
at the Hamiltonian level and may only be broken spontaneously.

When the electric displacement field is large, we project $H$ onto the lowest conduction band and solve it self-consistently within Hartree--Fock theory,
allowing both flavor polarization and translational symmetry breaking. The
crystal Bravais lattice is taken to be triangular with one electron per unit cell. Thus, the lattice constant $L_s$ is fixed by the density $L_s = \sqrt{2/(\sqrt{3}n)}$. The pinning potential with the leading order can be chosen as 
\begin{equation}
    V(\mathbf{r})=\sum_\bG \Delta(\bG) e^{i\bG \cdot \mathbf{r}},
\end{equation}
where only the first shell of the reciprocal lattice vector $\bG$ is preserved. This potential acts on the bottom layer of RMG. The phase diagram for R5G with no pinning potential is shown in Fig.~1. At low density, the
system spontaneously polarizes into a single spin-valley flavor. Within this
flavor-polarized sector, we find a competition between a quarter metal (QM), a
topologically trivial Wigner crystal (WC), an anomalous Hall crystal (AHC), and
a halo anomalous Hall crystal (hAHC). The phase boundaries are first order as they arise from level crossings between distinct mean-field solutions.

\begin{figure*}
\includegraphics[width=2\columnwidth]{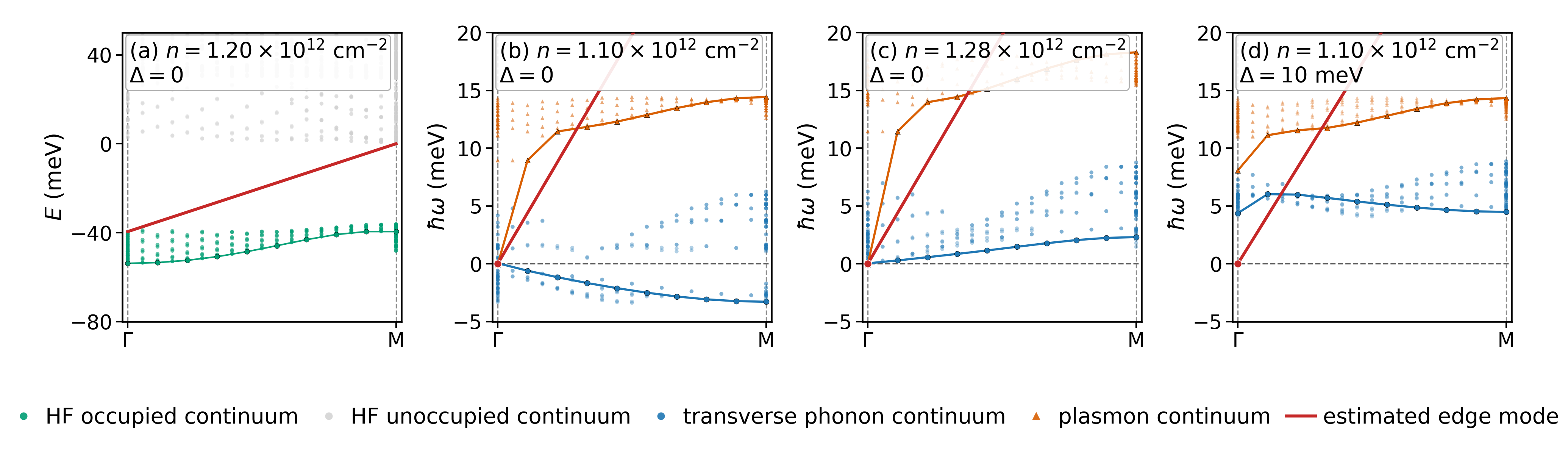}
\caption{
(a) Hartree--Fock energies of AHC projected along the $\Gamma$--$M$ direction. The red line schematically indicates the estimated chiral edge-mode dispersion connecting the occupied (green dots) and unoccupied states (gray dots). 
(b) Boundary-projected TDHF collective-mode spectrum along the same direction, shown near the continuous transition. The transverse phonon continuum overlaps with the chiral edge mode, enabling bulk--edge hybridization. (c),(d) Same as (b), but for varying electron densities $n$ and pinning potentials $\Delta$.}
\label{fig3}
\end{figure*}

For interlayer potential $U\simeq 30$--$50$ meV, increasing density produces the sequence: ${\rm WC} \rightarrow {\rm AHC} \rightarrow {\rm QM}$.
Thus the AHC appears as an intermediate quantum crystalline phase between the
classical Wigner crystal and the flavor-polarized metal. It retains charge
crystalline order, but its reconstructed bands acquire a nonzero Chern number
through band inversion at the mini-Brillouin-zone $K$ and $K'$ points
\cite{valentin25efficient, miao2026ahc}. At still larger displacement fields, the low-density WC regime is replaced by
metallic or hAHC states in the density window
$n \sim (0.4$--$0.8)\times 10^{12}\,{\rm cm}^{-2}$. This trend is associated
with a change in the single-particle band structure: the lowest conduction band
evolves from a flat-bottomed band with strong van Hove enhancement to a more
pronounced Mexican-hat-like dispersion, which suppresses the density of states
near the relevant filling \cite{bernevig_berry_2025,miao2026ahc}. The hAHC
differs from the AHC in its band-inversion mechanism: the inversion occurs near
$\Gamma$ in the hAHC, whereas it occurs near the mini-Brillouin-zone $K$ and
$K'$ points in the AHC. This difference in the Bloch band structure is closely related to recent discussion on the self doped metallic Wigner crystal \cite{metallic_wc_kim, feng26metalicWC,dong26metalicWC,soejima_ensuremathlambda-jellium_2025,han2026WC}. 

\textcolor{NavyBlue}{\textit{Softened phonons and instability.}}
The AHC breaks translational symmetry and supports
low-energy collective modes associated with distortions of the electronic
crystal. These collective modes can be computed by time-dependent Hartree-Fock approximation \cite{cote1991collective, ambuj2025phonon, kwan_mean-field_2025,dong_phonons_2025,shao25elecmag,mark2026phonon} which is expected to be reliable at long-wavelength since it restores the broken symmetry. In this approach, the particle-hole excitation spectrum is obtained by diagonalizing  the generalized-RPA (gRPA) matrix
\begin{equation}
R(\mathbf q)=
\begin{pmatrix}
A_{ph}(\mathbf q) & B_{ph}(\mathbf q)\\
-B^{*}_{ph}(-\mathbf q) & -A^{*}_{ph}(-\mathbf q)
\end{pmatrix},
\end{equation}
where \(p,h\) denote intra-flavor particle--hole labels in the
flavor-polarized AHC. The \(A\) and \(B\) blocks include both direct and exchange
scattering processes. The term ``generalized''  means we also account for exchange scattering, unlike the Bohm--Pines RPA for plasmon.

When an eigenfrequency becomes imaginary, the corresponding TDHF mode is
dynamically unstable. An imaginary frequency 
 leads to fluctuations whose amplitude grows exponentially in time, $\delta\rho_{ph}(t)= X_{ph} e^{i\omega t} +Y_{ph}^*e^{-i\omega^{*} t}$,
eventually invalidating the linearized approximation. These unstable regions are indicated by dashed lines in Fig.~\ref{fig1}.  It occurs mainly inside the AHC phase at large displacement field $D$. As shown in the SM \cite{Supp}, this is mainly due to the decrease of the density of states with increasing $D$, after a van-Hove singularity \cite{Desrohcers26elastic}.
We also find that an instability appears when the AHC approaches the (quarter) metal phase, suggesting the transition could be a continuous phase transition. 

A representative density-driven instability at fixed displacement field is
shown in Fig.~\ref{fig2}. As the AHC--WC boundary is approached, the phonon mode
at the \(M\) point of the reduced Brillouin zone softens and becomes imaginary,
with the unstable branch extending over a sizable portion of the
\(\Gamma\)--\(M\) line. TDHF study cannot determine the nature of the new ground state, but provides hints about which ground state is most likely to appear. Since the mode with the largest imaginary frequency appears at the $M$ point, the new ground state likely has the corresponding charge modulation. As shown in the SM \cite{Supp}, the charge density has a stripe-like feature. Such a stripe state could be related to a spin-valley polarized metallic state in rhombohedral hexalayer graphene which exhibits highly anisotropic transport behavior \cite{qin25stripe}.
Upon further increasing the carrier density, this imaginary mode hardens and
becomes real, signaling the restoration of dynamical stability. The softened mode in the TDHF spectrum \cite{kwan_moire_2025,dong_phonons_2025} can also be stabilized by a weak pinning potential $\Delta(\mathbf r)$ [Fig.~\ref{fig2}(e)--(h)], which pushes the low-lying phonon mode to higher energy and makes the system behave more like an ordinary Chern insulator \cite{Shi_fractional_2025}.

We now study the Hartree--Fock eigenvalues
$\epsilon_{\rm HF}(\vec{k})$ and the gRPA eigenvalues
$\omega_{\rm gRPA}(\vec{q})$ by projecting them onto high-symmetry directions
of the hexagonal Brillouin zone. To be concrete, we consider a cut from the center $\Gamma$ to the face of the hexagonal Brillouin zone $M$, which we define to be the $q_x$ direction.
Fig.~\ref{fig3}(a) shows the HF spectrum along this direction $\epsilon_{\rm HF}(q_x,k_y)$ for all
transverse momenta $k_y$.
This projected spectrum allows us to estimate the chiral edge-mode velocity, since the edge mode must
interpolate between the occupied (valence) states to the unoccupied (conduction) states
of the AHC.  Fig.~\ref{fig3}(b)--(d) show the
gRPA spectrum $\omega_{\rm gRPA}(\vec{q})$ projected along the same direction for
several densities, with/without an external periodic potential $\Delta$.
The blue (orange) points correspond to transverse (longitudinal) phonon modes. Near the continuous transition, where the phonon mode softens, the chiral edge
mode necessarily crosses the projected phonon continuum, as shown in Figs.~\ref{fig3}(b) and (c).

\begin{figure}
\includegraphics[width=0.95\columnwidth]{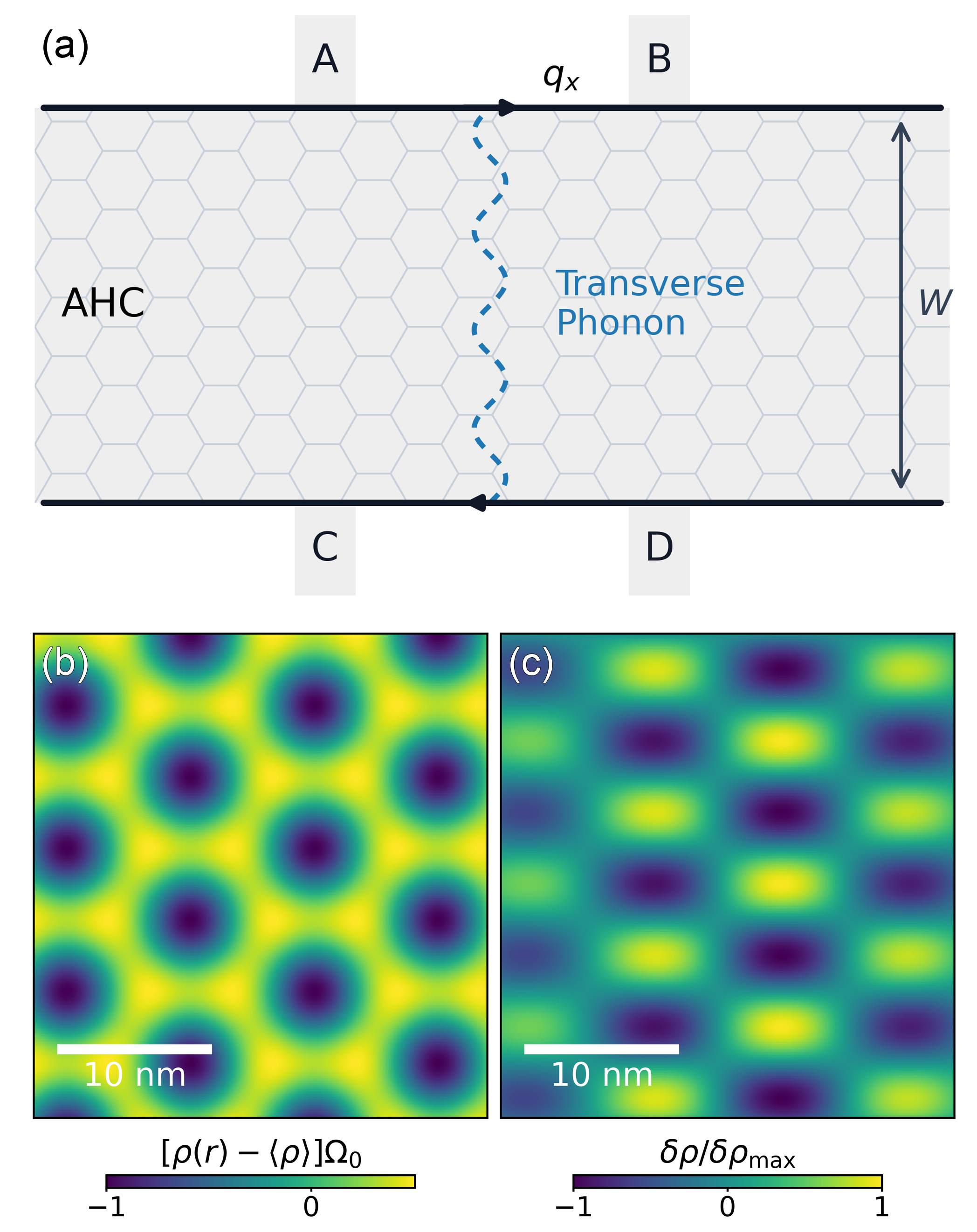}
\caption{(a) Experimental schematic of retarded inter-edge coupling in a four-terminal anomalous Hall crystal (AHC) device. A microwave drive applied between top contacts A and B excites a chiral edge wave, which resonantly emits a transverse bulk phonon. This phonon propagates across the sample width $W$ and induces a phase-delayed nonlocal response measured between bottom contacts C and D. The honeycomb pattern hints at the underlying AHC order. (b) Charge density profile $n(\mathbf{r})$ for the AHC ground state. The doping density is taken as $1.3 \times 10^{12} \,\rm cm^{-2}$. (c) Charge density modulation $\delta n_\bq(\mathbf{r})$ induced by a long-wavelength transverse phonon, $\bq$ is taken as $1/12(\bG_1+\bG_2)$, which is on the $\Gamma-M$ path. The corresponding $k$-space geometry is shown in SM \cite{Supp}.}
\label{fig:schematic}
\end{figure}

\textcolor{NavyBlue}{\textit{Bulk-Edge Coupling.}}
We now consider a strip geometry for AHC with bottom ($\rm B$) and top ($\rm T$) edges at $y_{\rm B}=0$ and $y_{\rm T}=W$, shown in Fig.~\ref{fig:schematic}(a). The transverse bulk mode couples naturally to the edge density $\rho_\eta(x,t)$ through the shear strain \cite{heinonen96epchall},
\begin{equation}
S_{\rm int}
=
g
\sum_{\eta={\rm B,T}}
\int dt dx\,
\rho_\eta(x,t)\,
u_{xy}(x,y_\eta,t),
\end{equation}
where $u_{xy}=\frac{1}{2}(\partial_x u_y+\partial_y u_x),
$ and $\mathbf u$ is the displacement field. Importantly, the coupling constant $g$ between the low-lying bulk collective mode and the edge mode is symmetry-allowed, and can originate from the Coulomb potential. We provide an estimation for $g$ in SM \cite{Supp}. This is qualitatively different from valley-polarized Chern insulators \cite{eslam20softmode, yves21exciton, wang23magorder, qiu25magnon, ming26spin}, where the bulk collective modes are inter-flavor excitations and are hence decoupled from the (intra-flavor) edge mode. We note that $g$ is distinct from a conventional electron--phonon coupling \cite{giustino2017electron} because both the transverse bulk collective mode and the chiral edge mode of the AHC originate from the same itinerant electronic degrees of freedom in the continuum model described by Eq.~1. The ``lattice'' displacement of the AHC is therefore not an independent ionic coordinate, but a collective vibration of the electronic charge texture itself [Fig.~\ref{fig:schematic}(b), (c)].
Integrating out the bulk displacement field, as shown in the SM \cite{Supp}, gives the induced interaction between edge densities,
\begin{equation}
S_{\rm ind}
=
-
\int\frac{d\omega dq_x}{8\pi^2}
\rho_\eta(-q_x,-\omega)
U_{\eta\eta'}(q_x,\omega)
\rho_{\eta'}(q_x,\omega),
\end{equation}
\begin{align}
U_{\eta\eta'}(q_x,\omega)
=
g^2
\int& \frac{dq_y}{2\pi} \big[
C_i(\mathbf q)
D_{ij}^R(q_x,q_y,\omega)
C_j^*(\mathbf q)\nonumber \\
&e^{iq_y(y_\eta-y_{\eta'})}\big].
\end{align}
Here $C_i(\mathbf q)=\frac{\rm i}{2}(q_y,q_x)
$ is the shear-strain vertex, $D_{ij}^R$ is the retarded bulk phonon propagator, and repeated indices ($i,j\in(x,y)$) are summed over. Since the intra-edge interaction is expected to be dominated by the instantaneous Coulomb interaction, we focus on the inter-edge component,
\begin{equation}
U_{\rm BT}^R(q_x,\omega)
=
g^2
\int\frac{dq_y}{2\pi}
C_i(\mathbf q)
D_{ij}^R(q_x,q_y,\omega)
C_j^*(\mathbf q)
e^{-iq_y W}.
\end{equation}
This integral is controlled by the analytic structure of the bulk propagator in the complex $q_y$ plane. For a given external frequency $\omega$ and longitudinal momentum $q_x$, the relevant poles are determined by the solutions of
\begin{equation}
    \omega=\omega_{\rm gRPA}(q_x,k_y),
\end{equation}
where $k_y$ is allowed to be complex. In general, this equation should be solved for all modes of the gRPA spectrum. In the low-frequency regime $\omega\ll \Delta_{\rm HF}$, the Hartree-Fock energy gap, however, only the low-lying collective modes are relevant, namely the plasmon and the transverse phonon mode. If the relevant pole has a nonzero imaginary part, ${\rm Im}\,k_y=\kappa\neq 0$, the bulk response is evanescent across the strip and the inter-edge interaction is exponentially suppressed $U_{\rm BT}^R\propto e^{-\kappa W}$.
However, if the pole occurs at real $k_y$, the bulk collective mode can propagate across the strip without exponentially suppressed by the sample width $W$.
For illustration, let us take an isotropic sound mode with dispersion $\omega_T(q)=c_Tq$, 
\begin{equation}
D_{ij}^{R}(q,\omega)
=
\frac{\delta_{ij}-q_iq_j/q^2}
{\rho_M[(\omega+i0^+)^2-c_T^2(q_x^2+q_y^2)]}.
\end{equation}
Here $\rho_M$ is the inertia density of the transverse sound and $c_T$ is its velocity. We can satisfy energy-momentum conservation only for $\omega>c_T|q_x|$, then,
$k_y(q,\omega)=
\sqrt{\frac{\omega^2}{c_T^2}-q_x^2}$. Evaluating the $q_y$ integral by closing the contour around this pole gives the following,
\begin{equation}
U_{\rm BT}^R(q_x,\omega)
=
-i
\frac{g^2}
{2\rho_M c_T^2 k_y}
\left[
\frac{1}{4}
\frac{(q_x^2-k_y^2)^2}{q_x^2+k_y^2}
\right]
e^{ik_yW}.
\end{equation}
This interaction takes a simple form in the local limit:
\begin{equation}
U_{\rm BT}^R(0,\omega)
=
-\frac{i g^2}{8\rho_M c_T^3}
\,\omega\,e^{i\omega W/c_T}
\end{equation}
The phase factor $e^{i\omega W/c_T}$ is the phase accumulated by the transverse wave as it travels across the sample width $W$ with speed $c_T$.
Fourier transforming back to the time domain, the induced interaction is local along the edge but delayed by the transverse-mode time of flight, $W/c_T$:
\begin{equation}
S_{\rm ind}^{{\rm TB}}
=
-\frac{g^2}{8\rho_M c_T^3}
\int dx\,dt\,
\rho_{\rm B}(x,t)\,
\partial_t
\rho_{\rm T}\left(x,t-\frac{W}{c_T}\right),
\end{equation} 

The resulting edge theory is thus a Luttinger-liquid variant with retarded inter-edge interactions, while intra-edge interactions are instantaneous. 

The above interaction has an important consequence. Suppose an edge excitation is created with dispersion $\omega_{\rm edge}(q_x)=v q_x$. When the edge velocity is smaller than the transverse sound velocity, $v<c_T$, the edge excitation is below the bulk transverse-mode continuum. Then, the bulk-mediated coupling to the opposite edge is exponentially suppressed by $W$. However, when $v>c_T$, the edge mode can resonantly emit a propagating transverse collective mode into the bulk since there is always a real $k_y$ that satisfies the energy longitudinal-momentum conservation $vq_x=c_T\sqrt{q_x^2+k_y^2}$.
Such an energy--momentum matching condition also appears when launching magnons from a $\nu=1$ region into a $\nu=0$ region across a quantum Hall ferromagnet junction \cite{wei2021scattering}.

\textcolor{NavyBlue}{\textit{Summary.}} 
We carried out a detailed TDHF study of R5G and found that the transverse sound mode of the anomalous Hall crystal (AHC) is generically very soft, suggesting a proximate instability toward a stripe phase. This soft bulk mode has an important dynamical consequence: an edge excitation at the boundary of the AHC can radiate a propagating transverse collective mode when its velocity exceeds the bulk transverse sound velocity.
Experimental identification of this phenomena would provide a smoking gun signature of AHC.

\textcolor{NavyBlue}{\textit{Acknowledgement.}} We thank Daniel Parker, Yves Kwan, Yuelin Shao,  Nemin Wei and Yongxin Zheng for insightful discussions. W.M, M.C. and B.Y. acknowledge the financial support by the Penn State Materials Research Science and Engineering Center for Nanoscale Science (MRSEC) under National Science Foundation award DMR-2011839. C.~Huang is supported by the U.S. Department of Energy, Office of Science, Office of Basic Energy Sciences, under Award Number DE-SC-0024346.

\bibliographystyle{apsrev4-2}
\bibliography{reference}

@article{giustino2017electron,
	title        = {Electron-phonon interactions from first principles},
	author       = {Giustino, Feliciano},
	year         = 2017,
	journal      = {Reviews of Modern Physics},
	publisher    = {APS},
	volume       = 89,
	number       = 1,
	pages        = {015003}
}

@article{wei2021scattering,
	title        = {Scattering of magnons at graphene quantum-Hall-magnet junctions},
	author       = {Wei, Nemin and Huang, Chunli and MacDonald, Allan H},
	year         = 2021,
	journal      = {Physical Review Letters},
	publisher    = {APS},
	volume       = 126,
	number       = 11,
	pages        = 117203
}

@article{cote1991collective,
	title        = {Collective modes of the two-dimensional Wigner crystal in a strong magnetic field},
	author       = {C{\^o}t{\'e}, Ren{\'e} and MacDonald, AH},
	year         = 1991,
	journal      = {Physical Review B},
	publisher    = {APS},
	volume       = 44,
	number       = 16,
	pages        = 8759
}

@article{chang2023colloquium,
	title        = {Colloquium: quantum anomalous Hall effect},
	author       = {Chang, Cui-Zu and Liu, Chao-Xing and MacDonald, Allan H},
	year         = 2023,
	journal      = {Reviews of Modern Physics},
	publisher    = {APS},
	volume       = 95,
	number       = 1,
	pages        = {011002}
}

@article{dong_anomalous_2024,
	title        = {Anomalous {Hall} {Crystals} in {Rhombohedral} {Multilayer} {Graphene}. {I}. {Interaction}-{Driven} {Chern} {Bands} and {Fractional} {Quantum} {Hall} {States} at {Zero} {Magnetic} {Field}},
	author       = {Dong, Junkai},
	year         = 2024,
	journal      = {Physical Review Letters},
	volume       = 133,
	number       = 20
}

@article{soejima_ensuremathlambda-jellium_2025,
	title        = {\${\textbackslash}ensuremath\{{\textbackslash}lambda\}\$-{Jellium} {Model} for the {Anomalous} {Hall} {Crystal}},
	author       = {Soejima, Tomohiro and Dong, Junkai and Vishwanath, Ashvin and Parker, Daniel E.},
	year         = 2025,
	month        = oct,
	journal      = {Physical Review Letters},
	volume       = 135,
	number       = 18,
	pages        = 186505,
	note         = {Publisher: American Physical Society}
}

@article{dong_phonons_2025,
	title        = {Phonons in electron crystals with {Berry} curvature},
	author       = {Dong, Junkai and Sommer, Ophelia Evelyn and Soejima, Tomohiro and Parker, Daniel E. and Vishwanath, Ashvin},
	year         = 2025,
	month        = sep,
	journal      = {Proceedings of the National Academy of Sciences},
	volume       = 122,
	number       = 38,
	pages        = {e2515532122},
	note         = {Publisher: Proceedings of the National Academy of Sciences}
}

@article{soejima_anomalous_2024,
	title        = {Anomalous {Hall} crystals in rhombohedral multilayer graphene. {II}. {General} mechanism and a minimal model},
	author       = {Soejima, Tomohiro and Dong, Junkai and Wang, Taige and Wang, Tianle and Zaletel, Michael P. and Vishwanath, Ashvin and Parker, Daniel E.},
	year         = 2024,
	month        = nov,
	journal      = {Physical Review B},
	volume       = 110,
	number       = 20,
	pages        = 205124,
	note         = {Publisher: American Physical Society}
}

@article{dong_stability_2024,
	title        = {Stability of anomalous {Hall} crystals in multilayer rhombohedral graphene},
	author       = {Dong, Zhihuan and Patri, Adarsh S. and Senthil, T.},
	year         = 2024,
	month        = nov,
	journal      = {Physical Review B},
	volume       = 110,
	number       = 20,
	pages        = 205130,
	note         = {Publisher: American Physical Society}
}

@article{dong_theory_2024,
	title        = {Theory of {Quantum} {Anomalous} {Hall} {Phases} in {Pentalayer} {Rhombohedral} {Graphene} {Moir}{\textbackslash}'e {Structures}},
	author       = {Dong, Zhihuan and Patri, Adarsh S. and Senthil, T.},
	year         = 2024,
	month        = nov,
	journal      = {Physical Review Letters},
	volume       = 133,
	number       = 20,
	pages        = 206502,
	note         = {Publisher: American Physical Society}
}

@misc{tan_ideal_2025,
	title        = {The ideal limit of rhombohedral graphene: {Interaction}-induced layer-skyrmion lattices and their collective excitations},
	shorttitle   = {The ideal limit of rhombohedral graphene},
	author       = {Tan, Tixuan and Ledwith, Patrick J. and Devakul, Trithep},
	year         = 2025,
	month        = nov,
	publisher    = {arXiv},
	note         = {arXiv:2511.07402 [cond-mat]}
}

@article{tan_parent_2024,
	title        = {Parent {Berry} {Curvature} and the {Ideal} {Anomalous} {Hall} {Crystal}},
	author       = {Tan, Tixuan and Devakul, Trithep},
	year         = 2024,
	month        = nov,
	journal      = {Physical Review X},
	volume       = 14,
	number       = 4,
	pages        = {041040},
	note         = {Publisher: American Physical Society}
}

@article{zhou_new_2025,
	title        = {New {Classes} of {Quantum} {Anomalous} {Hall} {Crystals} in {Multilayer} {Graphene}},
	author       = {Zhou, Boran and Zhang, Ya-Hui},
	year         = 2025,
	month        = jul,
	journal      = {Physical Review Letters},
	volume       = 135,
	number       = 3,
	pages        = {036501},
	note         = {Publisher: American Physical Society}
}

@misc{kwan_mean-field_2025,
	title        = {Mean-field {Modelling} of {Moiré} {Materials}: {A} {User}'s {Guide} with {Selected} {Applications} to {Twisted} {Bilayer} {Graphene}},
	shorttitle   = {Mean-field {Modelling} of {Moiré} {Materials}},
	author       = {Kwan, Yves H. and Wang, Ziwei and Wagner, Glenn and Bultinck, Nick and Simon, Steven H. and Parameswaran, Siddharth A.},
	year         = 2025,
	month        = nov,
	publisher    = {arXiv},
	note         = {arXiv:2511.21683 [cond-mat]}
}

@article{kwan_moire_2025,
	title        = {Moir{\textbackslash}'e fractional {Chern} insulators. {III}. {Hartree}-{Fock} phase diagram, magic angle regime for {Chern} insulator states, role of moir{\textbackslash}'e potential, and {Goldstone} gaps in rhombohedral graphene superlattices},
	author       = {Kwan, Yves H. and Yu, Jiabin and Herzog-Arbeitman, Jonah and Efetov, Dmitri K. and Regnault, Nicolas and Bernevig, B. Andrei},
	year         = 2025,
	month        = aug,
	journal      = {Physical Review B},
	volume       = 112,
	number       = 7,
	pages        = {075109},
	note         = {Publisher: American Physical Society}
}

@misc{bernevig_berry_2025,
	title        = {"{Berry} {Trashcan}" {Model} of {Interacting} {Electrons} in {Rhombohedral} {Graphene}},
	author       = {Bernevig, B. Andrei and Kwan, Yves H.},
	year         = 2025,
	month        = mar,
	publisher    = {arXiv},
	note         = {arXiv:2503.09692 [cond-mat]}
}

@article{herzog-arbeitman_moire_2024,
	title        = {Moir{\textbackslash}'e fractional {Chern} insulators. {II}. {First}-principles calculations and continuum models of rhombohedral graphene superlattices},
	author       = {Herzog-Arbeitman, Jonah and Wang, Yuzhi and Liu, Jiaxuan and Tam, Pok Man and Qi, Ziyue and Jia, Yujin and Efetov, Dmitri K. and Vafek, Oskar and Regnault, Nicolas and Weng, Hongming and Wu, Quansheng and Bernevig, B. Andrei and Yu, Jiabin},
	year         = 2024,
	month        = may,
	journal      = {Physical Review B},
	volume       = 109,
	number       = 20,
	pages        = 205122,
	note         = {Publisher: American Physical Society}
}

@article{zeng_sublattice_2024,
	title        = {Sublattice {Structure} and {Topology} in {Spontaneously} {Crystallized} {Electronic} {States}},
	author       = {Zeng, Yongxin and Guerci, Daniele and Crépel, Valentin and Millis, Andrew J. and Cano, Jennifer},
	year         = 2024,
	month        = jun,
	journal      = {Physical Review Letters},
	volume       = 132,
	number       = 23,
	pages        = 236601,
	note         = {Publisher: American Physical Society}
}

@article{lu_fractional_2024,
	title        = {Fractional quantum anomalous {Hall} effect in multilayer graphene},
	author       = {Lu, Zhengguang and Han, Tonghang and Yao, Yuxuan and Reddy, Aidan P. and Yang, Jixiang and Seo, Junseok and Watanabe, Kenji and Taniguchi, Takashi and Fu, Liang and Ju, Long},
	year         = 2024,
	month        = feb,
	journal      = {Nature},
	volume       = 626,
	number       = 8000,
	pages        = {759--764},
	copyright    = {2024 The Author(s), under exclusive licence to Springer Nature Limited},
	note         = {Publisher: Nature Publishing Group}
}

@article{lu_extended_2025,
	title        = {Extended quantum anomalous {Hall} states in graphene/{hBN} moiré superlattices},
	author       = {Lu, Zhengguang and Han, Tonghang and Yao, Yuxuan and Hadjri, Zach and Yang, Jixiang and Seo, Junseok and Shi, Lihan and Ye, Shenyong and Watanabe, Kenji and Taniguchi, Takashi and Ju, Long},
	year         = 2025,
	month        = jan,
	journal      = {Nature},
	volume       = 637,
	number       = 8048,
	pages        = {1090--1095},
	copyright    = {2025 The Author(s), under exclusive licence to Springer Nature Limited},
	note         = {Publisher: Nature Publishing Group}
}

@article{aronson_displacement_2025,
	title        = {Displacement {Field}-{Controlled} {Fractional} {Chern} {Insulators} and {Charge} {Density} {Waves} in a {Graphene}/{hBN} {Moiré} {Superlattice}},
	author       = {Aronson, Samuel H. and Han, Tonghang and Lu, Zhengguang and Yao, Yuxuan and Butler, Jackson P. and Watanabe, Kenji and Taniguchi, Takashi and Ju, Long and Ashoori, Raymond C.},
	year         = 2025,
	month        = jul,
	journal      = {Physical Review X},
	volume       = 15,
	number       = 3,
	pages        = {031026}
}

@article{Ghorashi_topological_2023_prl,
	title        = {Topological and Stacked Flat Bands in Bilayer Graphene with a Superlattice Potential},
	author       = {Ghorashi, Sayed Ali Akbar and Dunbrack, Aaron and Abouelkomsan, Ahmed and Sun, Jiacheng and Du, Xu and Cano, Jennifer},
	year         = 2023,
	month        = {May},
	journal      = {Phys. Rev. Lett.},
	publisher    = {American Physical Society},
	volume       = 130,
	pages        = 196201,
	issue        = 19,
	numpages     = 8
}

@article{Miao_artificial_2025_prb,
	title        = {Artificial moir\'e engineering for an ideal Bernevig-Hughes-Zhang model},
	author       = {Miao, Wangqian and Rashidi, Arman and Dai, Xi},
	year         = 2025,
	month        = {Jan},
	journal      = {Phys. Rev. B},
	publisher    = {American Physical Society},
	volume       = 111,
	pages        = {045113},
	issue        = 4,
	numpages     = 9
}

@article{Shi_fractional_2025,
	title        = {Fractional Topological States in Rhombohedral Multilayer Graphene Modulated by Kagome Superlattice},
	author       = {Shi, Yanran and Xie, Bo and Ren, Fengfan and Cai, Xinyu and Guo, Zhongqing and Li, Qiao and Lu, Xin and Regnault, Nicolas and Liu, Zhongkai and Liu, Jianpeng},
	year         = 2025,
	month        = {Dec},
	journal      = {Phys. Rev. Lett.},
	publisher    = {American Physical Society},
	volume       = 135,
	pages        = 256603,
	issue        = 25,
	numpages     = 7
}

@article{tesanovic_hall_1989_prb,
	title        = {``Hall crystal'' versus Wigner crystal},
	author       = {Te\ifmmode \check{s}\else \v{s}\fi{}anovi\ifmmode \acute{c}\else \'{c}\fi{}, Zlatko and Axel, Fran\ifmmode \mbox{\c{c}}\else \c{c}\fi{}oise and Halperin, B. I.},
	year         = 1989,
	month        = {Apr},
	journal      = {Phys. Rev. B},
	publisher    = {American Physical Society},
	volume       = 39,
	pages        = {8525--8551},
	issue        = 12,
	numpages     = {0}
}

@misc{Huo_does_2025,
	title        = {Does Moire Matter? Critical Moire Dependence with Quantum Fluctuations in Graphene Based Integer and Fractional Chern Insulators},
	author       = {Zihao Huo and Wenxuan Wang and Jian Xie and Yves H. Kwan and Jonah Herzog-Arbeitman and Zaizhe Zhang and Qiu Yang and Min Wu and Kenji Watanabe and Takashi Taniguchi and Kaihui Liu and Nicolas Regnault and B. Andrei Bernevig and Xiaobo Lu},
	year         = 2025,
	eprint       = {arXiv:2510.15309}
}

@misc{ambuj2025phonon,
	title        = {Elementary Excitations, Melting Temperature and Correlation Energy in Wigner Crystal},
	author       = {Ambuj Jain and Chunli Huang},
	year         = 2025,
	eprint       = {arXiv:2504.04538}
}

@misc{mark2026phonon,
	title        = {Topological phonons in anomalous Hall crystals},
	author       = {Mark R. Hirsbrunner and Félix Desrochers and Joe Huxford and Yong Baek Kim},
	year         = 2026,
	eprint       = {arXiv:2601.06246}
}

@article{tomohiro2025jellium,
	title        = {$\ensuremath{\lambda}$-Jellium Model for the Anomalous Hall Crystal},
	author       = {Soejima, Tomohiro and Dong, Junkai and Vishwanath, Ashvin and Parker, Daniel E.},
	year         = 2025,
	month        = {Oct},
	journal      = {Phys. Rev. Lett.},
	publisher    = {American Physical Society},
	volume       = 135,
	pages        = 186505,
	issue        = 18,
	numpages     = 7
}

@article{uchida2026nonabilien,
	title        = {Non-Abelian Chern Band in Rhombohedral Graphene Multilayers},
	author       = {Uchida, Taketo and Kawakami, Takuto and Koshino, Mikito},
	year         = 2026,
	month        = {Apr},
	journal      = {Phys. Rev. Lett.},
	publisher    = {American Physical Society},
	volume       = 136,
	pages        = 156602,
	issue        = 15,
	numpages     = 8
}

@article{guo2025ahcblg,
	title        = {Correlation stabilized anomalous Hall crystal in bilayer graphene},
	author       = {Guo,  Zhongqing and Liu,  Jianpeng},
	year         = 2025,
	month        = dec,
	journal      = {Nature Communications},
	publisher    = {Springer Science and Business Media LLC},
	volume       = 16,
	number       = 1
}

@article{miao2026ahc,
	title        = {Various electronic crystal phases in rhombohedral graphene multilayers},
	author       = {Miao, Wangqian and Li, Chu},
	year         = 2026,
	month        = {Apr},
	journal      = {Phys. Rev. B},
	publisher    = {American Physical Society},
	volume       = 113,
	pages        = 155136,
	issue        = 15,
	numpages     = 11
}

@misc{han2026WC,
	title        = {Evidence of Metallic Wigner Crystal in Rhombohedral Graphene},
	author       = {Tonghang Han and Jackson P. Butler and Shenyong Ye and Zhenqi Hua and Surajit Dutta and Zach Hadjri and Zhenghan Wu and Jixiang Yang and Junseok Seo and Phatthanon Pattanakanvijit and Emily Aitken and Kenji Watanabe and Takashi Taniguchi and Peng Xiong and Eli Zeldov and Zhengguang Lu and Raymond Ashoori and Long Ju},
	year         = 2026,
	eprint       = {arXiv:2604.00113}
}

@misc{dong26metalicWC,
	title        = {Crystals Caught Doping: Metallic Wigner Crystals in Rhombohedral Graphene},
	author       = {Junkai Dong and Tomohiro Soejima and Daniel E. Parker and Ashvin Vishwanath},
	year         = 2026,
	eprint       = {arXiv:2604.00114}
}

@misc{feng26metalicWC,
	title        = {Self-doped Crystal from Preempted Band-inversion Transitions},
	author       = {Jiechao Feng and Zhaoyu Han and Michael P. Zaletel and Zhihuan Dong},
	year         = 2026,
	eprint       = {arXiv:2604.09820}
}

@article{Desrohcers26lamdaN,
	title        = {Electronic crystal phases in the presence of nonuniform Berry curvature and tunable Berry flux: The ${\ensuremath{\lambda}}_{N}$-jellium model},
	author       = {Desrochers, F\'elix and Huxford, Joe and Hirsbrunner, Mark R. and Kim, Yong Baek},
	year         = 2026,
	month        = {Jan},
	journal      = {Phys. Rev. B},
	publisher    = {American Physical Society},
	volume       = 113,
	pages        = {045148},
	issue        = 4,
	numpages     = 32
}

@article{Desrohcers26elastic,
	title        = {Elastic Response and Instabilities of Anomalous Hall Crystals},
	author       = {Desrochers, F\'elix and Hirsbrunner, Mark R. and Huxford, Joe and Patri, Adarsh S. and Senthil, T. and Kim, Yong Baek},
	year         = 2026,
	month        = {Apr},
	journal      = {Phys. Rev. Lett.},
	publisher    = {American Physical Society},
	volume       = 136,
	pages        = 166503,
	issue        = 16,
	numpages     = 8
}

@article{zhou24moireless,
	title        = {Fractional Quantum Anomalous Hall Effect in Rhombohedral Multilayer Graphene in the Moir\'eless Limit},
	author       = {Zhou, Boran and Yang, Hui and Zhang, Ya-Hui},
	year         = 2024,
	month        = {Nov},
	journal      = {Phys. Rev. Lett.},
	publisher    = {American Physical Society},
	volume       = 133,
	pages        = 206504,
	issue        = 20,
	numpages     = 7
}

@article{thouless1960stability,
	title        = {Stability conditions and nuclear rotations in the Hartree-Fock theory},
	author       = {Thouless, David J},
	year         = 1960,
	journal      = {Nuclear Physics},
	publisher    = {Elsevier},
	volume       = 21,
	pages        = {225--232}
}

@article{thouless1962time,
	title        = {Time-dependent Hartree-Fock equations and rotational states of nuclei},
	author       = {Thouless, DJ and Valatin, JG},
	year         = 1962,
	journal      = {Nuclear Physics},
	publisher    = {Elsevier},
	volume       = 31,
	pages        = {211--230}
}

@article{valentin25efficient,
	title        = {Efficient Prediction of Superlattice and Anomalous Miniband Topology from Quantum Geometry},
	author       = {Cr\'epel, Valentin and Cano, Jennifer},
	year         = 2025,
	month        = {Jan},
	journal      = {Phys. Rev. X},
	publisher    = {American Physical Society},
	volume       = 15,
	pages        = {011004},
	issue        = 1,
	numpages     = 20
}

@misc{eslam20softmode,
	title        = {Soft modes in magic angle twisted bilayer graphene},
	author       = {Eslam Khalaf and Nick Bultinck and Ashvin Vishwanath and Michael P. Zaletel},
	year         = 2020,
	eprint       = {arXiv:2009.14827}
}

@misc{shao25elecmag,
	title        = {Electromagnetic responses of bilayer excitonic insulators: from exciton London equations to dipole and inverse dipole Hall effects},
	author       = {Yuelin Shao and Hao Shi and Xi Dai},
	year         = 2025,
	eprint       = {arXiv:2509.02142}
}

@misc{qin25stripe,
	title        = {Extreme Anisotropy in the Metallic and Superconducting Phases of Rhombohedral Hexalayer Graphene},
	author       = {Peiyu Qin and Hai-Tian Wu and Ron Q. Nguyen and Erin Morissette and Naiyuan J. Zhang and K. Watanabe and T. Taniguchi and J. I. A. Li},
	year         = 2025,
	eprint       = {arXiv:2504.05129}
}

@article{excitations_in_GWC,
	title        = {Mapping charge excitations in generalized Wigner crystals},
	author       = {Li, Hongyuan and Xiang, Ziyu and Regan, Emma and Zhao, Wenyu and Sailus, Renee and Banerjee, Rounak and Taniguchi, Takashi and Watanabe, Kenji and Tongay, Sefaattin and Zettl, Alex and Crommie, Michael F. and Wang, Feng},
	year         = 2024,
	journal      = {Nature Nanotechnology},
	volume       = 19,
	number       = 5,
	pages        = {618--623},
	isbn         = {1748-3395},
	date         = {2024/05/01},
	id           = {Li2024}
}

@article{2D_BWC_Goldoni_1996,
	title        = {Stability, dynamical properties, and melting of a classical bilayer Wigner crystal},
	author       = {Goldoni, G. and Peeters, F. M.},
	year         = 1996,
	month        = {Feb},
	journal      = {Phys. Rev. B},
	publisher    = {American Physical Society},
	volume       = 53,
	pages        = {4591--4603},
	issue        = 8,
	numpages     = {0}
}

@article{Wigner_original,
	title        = {On the Interaction of Electrons in Metals},
	author       = {Wigner, E.},
	year         = 1934,
	month        = {Dec},
	journal      = {Phys. Rev.},
	publisher    = {American Physical Society},
	volume       = 46,
	pages        = {1002--1011},
	issue        = 11,
	numpages     = {0}
}

@article{Needs_HFWC,
	title        = {Unrestricted Hartree-Fock theory of Wigner crystals},
	author       = {Trail, J. R. and Towler, M. D. and Needs, R. J.},
	year         = 2003,
	month        = {Jul},
	journal      = {Phys. Rev. B},
	publisher    = {American Physical Society},
	volume       = 68,
	pages        = {045107},
	issue        = 4,
	numpages     = 5
}

@article{Bernu_MIT_2008,
	title        = {Metal-insulator transition in the Hartree-Fock phase diagram of the fully polarized homogeneous electron gas in two dimensions},
	author       = {Bernu, B. and Delyon, F. and Duneau, M. and Holzmann, M.},
	year         = 2008,
	month        = {Dec},
	journal      = {Phys. Rev. B},
	publisher    = {American Physical Society},
	volume       = 78,
	pages        = 245110,
	issue        = 24,
	numpages     = 10
}

@article{sandeep25chiral,
	title        = {Chiral Wigner Crystal Phases Induced by Berry Curvature},
	author       = {Joy, Sandeep and Levitov, Leonid and Skinner, Brian},
	year         = 2025,
	month        = {Dec},
	journal      = {Phys. Rev. Lett.},
	publisher    = {American Physical Society},
	volume       = 135,
	pages        = 256502,
	issue        = 25,
	numpages     = 8
}

@misc{ming26spin,
	title        = {Collective Spin Excitations in Correlated Moiré Chern Ferromagnets},
	author       = {Ming Xie and Sankar Das Sarma},
	year         = 2026,
	eprint       = {arXiv:2603.20370}
}

@article{qiu25magnon,
	title        = {Topological magnons and domain walls in twisted bilayer ${\mathrm{MoTe}}_{2}$},
	author       = {Qiu, Wen-Xuan and Wu, Fengcheng},
	year         = 2025,
	month        = {Aug},
	journal      = {Phys. Rev. B},
	publisher    = {American Physical Society},
	volume       = 112,
	pages        = {085132},
	issue        = 8,
	numpages     = 14
}

@misc{wang23magorder,
	title        = {Diverse magnetic orders and quantum anomalous Hall effect in twisted bilayer MoTe2 and WSe2},
	author       = {Taige Wang and Trithep Devakul and Michael P. Zaletel and Liang Fu},
	year         = 2023,
	eprint       = {arXiv:2306.02501}
}

@article{yves21exciton,
	title        = {Exciton Band Topology in Spontaneous Quantum Anomalous Hall Insulators: Applications to Twisted Bilayer Graphene},
	author       = {Kwan, Yves H. and Hu, Yichen and Simon, Steven H. and Parameswaran, S. A.},
	year         = 2021,
	month        = {Mar},
	journal      = {Phys. Rev. Lett.},
	publisher    = {American Physical Society},
	volume       = 126,
	pages        = 137601,
	issue        = 13,
	numpages     = 5
}

@article{zeng25berryslide,
	title        = {Berry Phase Dynamics of Sliding Electron Crystals},
	author       = {Zeng, Yongxin and Millis, Andrew J.},
	year         = 2025,
	month        = {Aug},
	journal      = {Phys. Rev. X},
	publisher    = {American Physical Society},
	volume       = 15,
	pages        = {031059},
	issue        = 3,
	numpages     = 16
}

@article{patri24extended,
	title        = {Extended quantum anomalous Hall effect in moir\'e structures: Phase transitions and transport},
	author       = {Patri, Adarsh S. and Dong, Zhihuan and Senthil, T.},
	year         = 2024,
	month        = {Dec},
	journal      = {Phys. Rev. B},
	publisher    = {American Physical Society},
	volume       = 110,
	pages        = 245115,
	issue        = 24,
	numpages     = 10
}

@article{heinonen96epchall,
	title        = {Electron-Phonon Interactions on a Single-Branch Quantum Hall Edge},
	author       = {Heinonen, O. and Eggert, Sebastian},
	year         = 1996,
	month        = {Jul},
	journal      = {Phys. Rev. Lett.},
	publisher    = {American Physical Society},
	volume       = 77,
	pages        = {358--361},
	issue        = 2,
	numpages     = {0}
}

@misc{Qiao2026-iy,
Author = {Lei Qiao and Xin Lu and Fu-Chun Zhang and Jianpeng Liu},
Title = {Charge imprinting biases topology of correlated insulator in hBN-aligned rhombohedral multilayer graphene},
Year = {2026},
Eprint = {arXiv:2606.20377},
}

@misc{Luca_relaxation,
Author = {Luca Nashabeh and Héctor Ochoa},
Title = {Lattice Relaxation Flattens Chern Bands in Rhombohedral Graphene Stacks},
Year = {2026},
Eprint = {arXiv:2605.16218},
}

@misc{jacson_26_gaplessqah,
Author = {Jackson P. Butler and Tonghang Han and Andrew DiFabbio and Zach Hadjri and Emily Aitken and Kenji Watanabe and Takashi Taniguchi and Long Ju and Raymond C. Ashoori},
Title = {1/3 Fractional and Gapless Integer Quantum Anomalous Hall States in Rhombohedral Graphene},
Year = {2026},
Eprint = {arXiv:2606.06450},
}

@article{metallic_wc_kim,
  title = {Dynamical defects in a two-dimensional Wigner crystal: Self-doping and kinetic magnetism},
  author = {Kim, Kyung-Su and Esterlis, Ilya and Murthy, Chaitanya and Kivelson, Steven A.},
  journal = {Phys. Rev. B},
  volume = {109},
  issue = {23},
  pages = {235130},
  numpages = {13},
  year = {2024},
  month = {Jun},
  publisher = {American Physical Society}
}

@article{bfield_wc_kim,
  title = {Exchange interactions of a Wigner crystal in a magnetic field and Berry curvature: Multiparticle tunneling through complex trajectories},
  author = {Kim, Kyung-Su},
  journal = {Phys. Rev. B},
  volume = {113},
  issue = {14},
  pages = {144434},
  numpages = {11},
  year = {2026},
  month = {Apr},
  publisher = {American Physical Society}
}

@misc{Supp,
	title        = {Supplemental Material for ``Retarded Interaction Between Opposite Chiral Edges in Anomalous Hall Crystals''},
	year         = 2026,
	note         = {See Supplemental Material at [URL will be inserted by publisher]}
}

\onecolumngrid
\newpage

\begin{center}
  \textbf{\large Supplemental Material} \\[0.5em]
\end{center}
\setcounter{equation}{0}
\setcounter{figure}{0}
\setcounter{table}{0}
\setcounter{page}{1}
\setcounter{section}{0}

\renewcommand{\theequation}{S\arabic{equation}}
\renewcommand{\thefigure}{S\arabic{figure}}
\renewcommand{\thetable}{S\arabic{table}}
\renewcommand{\thesection}{S\Roman{section}}
\renewcommand{\thepage}{S\arabic{page}}

\section{Tight binding model for rhombohedral graphene}

To describe the low-energy electronic structure of rhombohedral multilayer graphene (RMG), we use the \emph{ab initio} Slater--Koster parametrization of the \(p_z\)-orbital tight-binding Hamiltonian developed in Ref.~\cite{herzog-arbeitman_moire_2024}. The hopping amplitude between two \(p_z\) orbitals separated by vector \(\mathbf{r}\), with \(r=|\mathbf{r}|\) and vertical projection \(z=\hat{z}\cdot \mathbf{r}\), is
\begin{equation}
    \begin{aligned}
        t_{\mathrm{SK}}(r)=&
        V_{pp\pi}\left(1-\frac{z^2}{r^2}\right)e^{q_\pi(1-r/a_\pi)}f_c(r)+
        V_{pp\sigma}\frac{z^2}{r^2}e^{q_\sigma(1-r/a_\sigma)}f_c(r),
    \end{aligned}
\end{equation}
where \(f_c(r)=\bigl(1+e^{(r-r_c)/l_c}\bigr)^{-1}\) is a smooth real-space cutoff. The optimized parameters are
\(V_{pp\pi}=-2.81~\mathrm{eV}\),
\(V_{pp\sigma}=0.48~\mathrm{eV}\),
\(a_\pi=1.418~\)\AA,
\(a_\sigma=3.349~\)\AA,
\(q_\pi=3.145\),
\(q_\sigma=7.428\),
\(r_c=6.14~\)\AA, and
\(l_c=0.265~\)\AA.
In addition to these hoppings, we include a small second-neighbor interlayer hopping \(t_2\approx -7~\mathrm{meV}\), which is important for reproducing the intrinsic gap of thin rhombohedral stacks. Two single-particle terms are further included to capture experimentally relevant symmetry breaking. The inversion-symmetric potential distinguishes the outer and inner layers,
\begin{equation}
    [H_{\mathrm{ISP}}]_{\ell\ell'}
    =
    V_{\mathrm{ISP}}\,
    \delta_{\ell\ell'}
    \left|
    \ell-\frac{N+1}{2}
    \right|,
\end{equation}
with \(V_{\mathrm{ISP}}/d_0\approx 5~\mathrm{meV}\)\AA. A perpendicular displacement field is modeled by
\begin{equation}
    [H_D]_{\ell\alpha,\ell'\beta}
    =
    U\left(\ell-\frac{N+1}{2}\right)
    \delta_{\ell\ell'}\delta_{\alpha\beta},
    \qquad
    U=\frac{|e|d_0D}{\varepsilon},
\end{equation}
where \(N\) is the number of graphene layers, \(d_0\) is the interlayer spacing, and \(\alpha,\beta\) label sublattice orbitals within each layer.

\begin{figure}
\includegraphics[width=0.55\columnwidth]{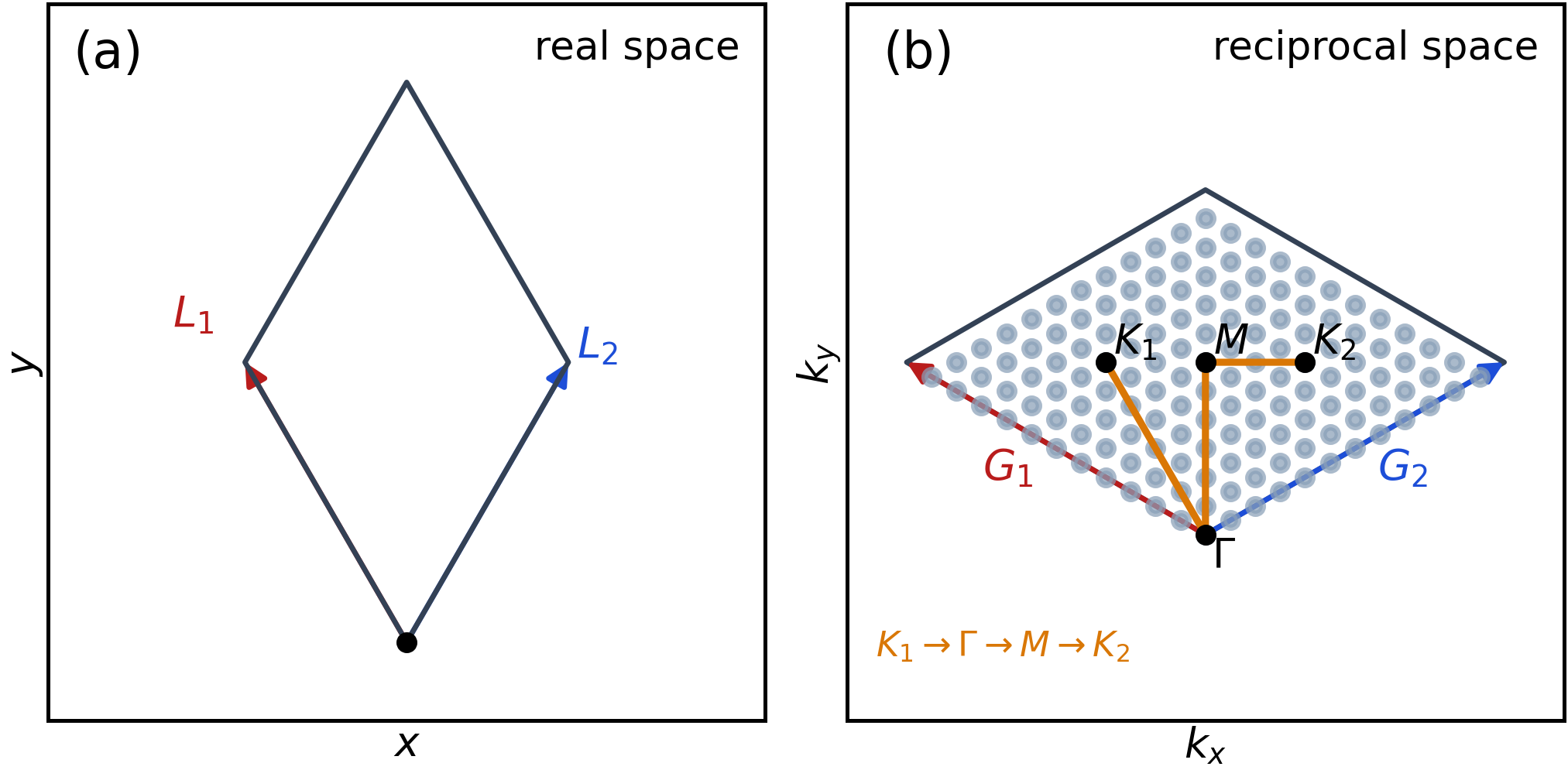}
\caption{
Schematic illustration of the real- and reciprocal-space conventions used in the calculations.
(a) Real-space moiré lattice generated by the primitive vectors $\mathbf{L}_1$ and $\mathbf{L}_2$.
(b) Reciprocal-space lattice generated by $\mathbf{G}_1$ and $\mathbf{G}_2$, with representative discrete momentum points and the high-symmetry path.
}
\label{figs0}
\end{figure}

\begin{figure}[!t]
\includegraphics[width=0.95\columnwidth]{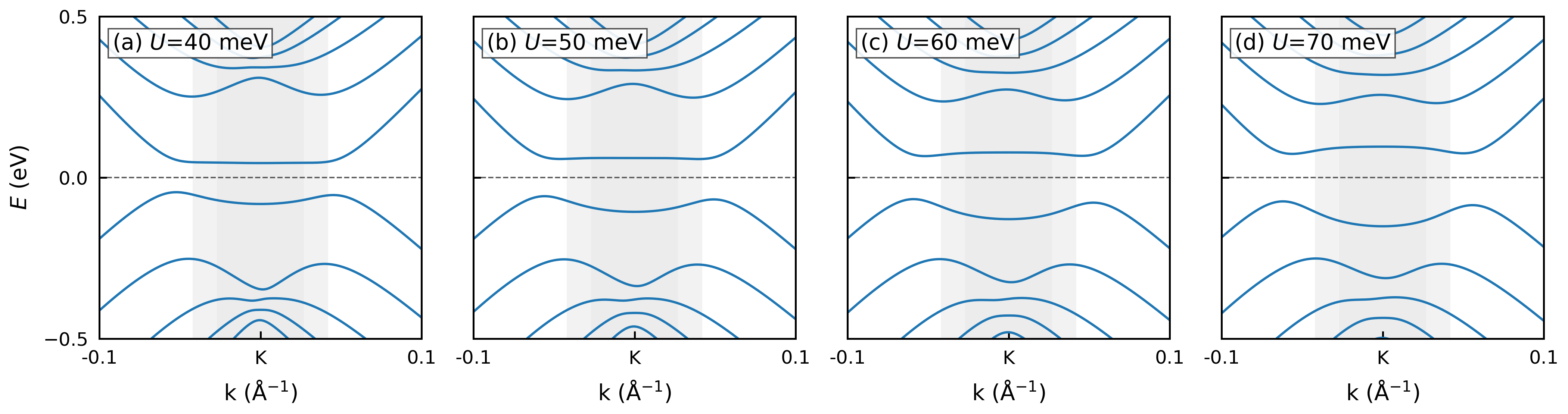}
\caption{
Single-particle band structures of R5G near $K$ valley for displacement potentials (a) $U$ = 40 meV, (b) $U$ = 50 meV, (c) $U$ = 60 meV, and (d) $U$ = 70 meV. Energies are measured relative to the charge-neutrality chemical potential. The shaded area labels the region when the electron filling is $0.5\times10^{12}$ cm$^{-2}$ and $1.2\times10^{12}$ cm$^{-2}$, respectively.
}
\label{figs1}
\end{figure}

For \(U>0\), the low-energy conduction-band states are predominantly localized on the outer-surface \(B_N\) orbital, while the valence-band states are mainly localized on the opposite-surface \(A_1\) orbital. The low-energy problem can therefore be reduced to an effective two-band description in the \(A_1\)--\(B_N\) surface subspace. In the bulk limit and at zero displacement field, these states connect continuously to the drumhead surface bands of rhombohedral graphite. At finite \(U\), the conduction-band minimum remains extremely flat, producing a large density of states near the band bottom. Furthermore, a stronger displacement field will make the band bottom dispersive. See Fig.~\ref{figs0} for the real space and reciprocal space convention used in this work and Fig.~\ref{figs1} for the corresponding band structures and we guide the readers to our previous work \cite{miao2026ahc} for more details of the density of states.

\section{Hartree Fock mean field formalism}

\begin{figure}
\includegraphics[width=0.9\columnwidth]{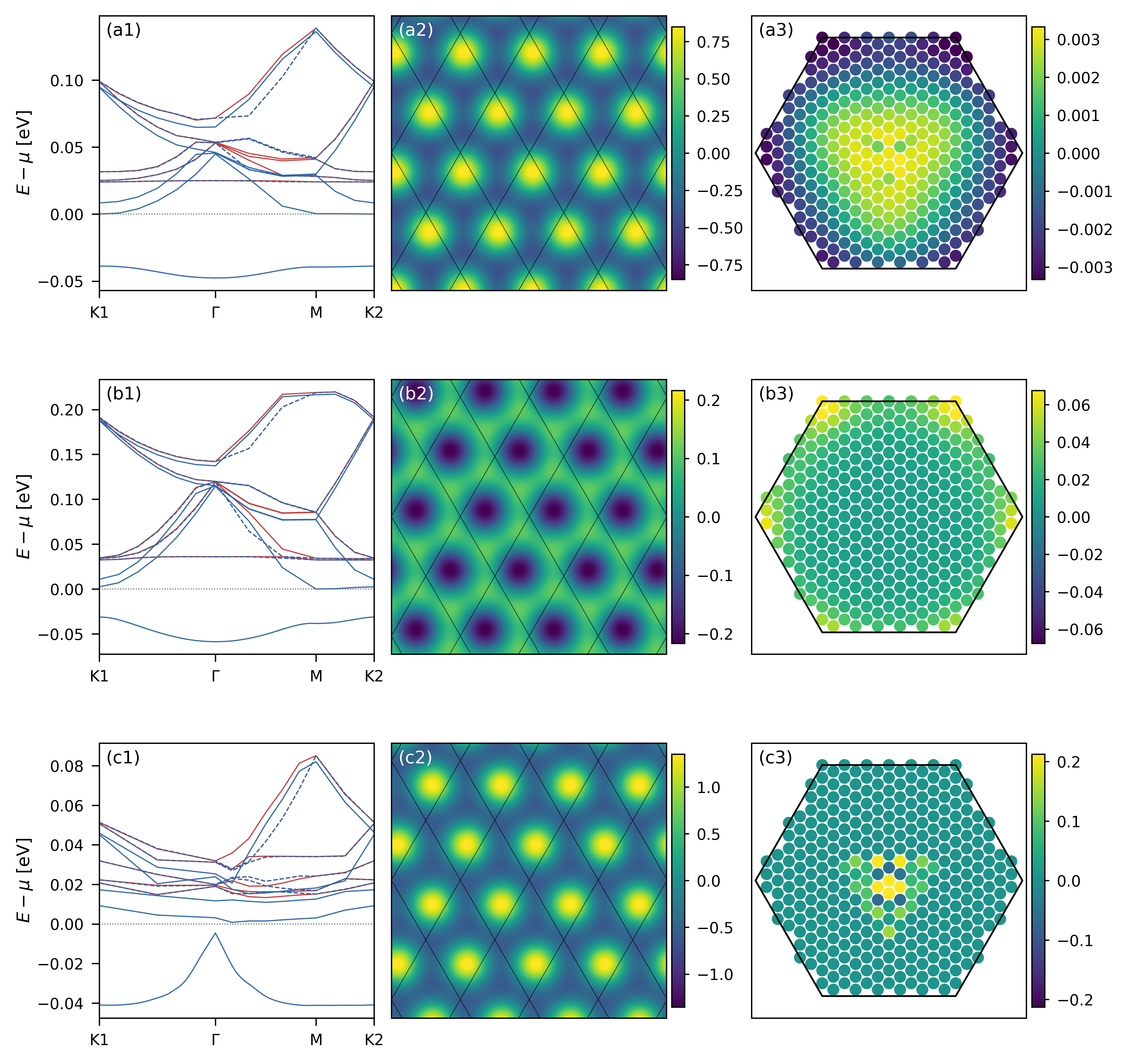}
\caption{
Hartree--Fock characterization of three representative ordered states in rhombohedral pentalayer graphene.
Rows show the WC, AHC, and hAHC states, respectively.
Columns show the Hartree--Fock band structure, the real-space charge density modulation
$(\rho(\mathbf r)-\langle \rho(\mathbf r)\rangle)\times \Omega_0$, and the Berry curvature distribution in the reduced Brillouin zone.
The parameters are $(C,n,U)=(0,0.8,40)$ for WC, $(1,1.4,50)$ for AHC, and $(1,0.8,65)$ for hAHC, where $C$ is the Chern number, $n$ is the doping density in units of $10^{12}\,\mathrm{cm}^{-2}$ and $U$ is in meV.
}
\label{figs2}
\end{figure}

To study the low-doping regime, we retain the long-range Coulomb interaction and neglect short-range interaction terms. The interacting Hamiltonian is written as
\begin{equation}
    H_{\mathrm{int}}
    =
    \frac{1}{2}
    \sum_{\bq}\sum_{\ell\ell'}
    v_{\ell\ell'}(\bq)\,
    :\rho_\ell(\bq)\rho_{\ell'}(-\bq):,
    \label{eq:Hint}
\end{equation}
with the layer-resolved density operator
\begin{equation}
    \rho_\ell(\bq)
    =
    \sum_{\bk,s,\mu,\sigma}
    c^\dagger_{\bk+\bq,s\mu\sigma\ell}\,
    c_{\bk,s\mu\sigma\ell}.
\end{equation}
Here \(s\) and \(\mu\) denote spin and valley, respectively, and \(\sigma\) labels the sublattice within layer \(\ell\). We use the screened interaction \cite{kwan_moire_2025}
\begin{equation}
    v_{\ell\ell'}(\bq)
    =
    \frac{e^2}{2S\varepsilon\varepsilon_0 q}
    \begin{cases}
        \tanh(qd_s), & \ell=\ell',\\
        e^{-q|\ell-\ell'|d_0}, & \ell\neq \ell',
    \end{cases}
\end{equation}
where \(S\) is the total area, \(d_s=10~\mathrm{nm}\) is the gate distance, \(d_0=0.335~\rm nm\) is the interlayer spacing of the graphene multilayer, and \(\varepsilon=5\) is the dielectric constant. Throughout the calculation we impose global neutrality by removing the uniform Hartree component, i.e. the \(q=0\) contribution is set to zero. For a chosen crystal geometry, the real-space superlattice constant is fixed by the carrier density \(n\): \(L_s=\sqrt{2/(\sqrt{3}n)}\) for a hexagonal lattice. We initialize the single-particle density matrix with different spin-valley occupations and iterate the Hartree--Fock Dyson equation to self-consistency. For each density and displacement field, the zero-temperature ground state is identified by comparing the converged condensation energies of the candidate metallic and symmetry-broken crystal states.

In practice, the interaction in Eq.~\eqref{eq:Hint} is projected onto an active low-energy subspace in the folded mini-Brillouin zone (mBZ). The projected interaction takes the form
\begin{equation}
    \begin{aligned}
        H_{\mathrm{int}}
        =
        \frac{1}{2N_{\bkbar}}
        \sum_{\bkbar,\bkbar',\bqbar}
        \sum_{\mu\mu'}\sum_{ss'}
        \sum_{nm\,n'm'}
        \left(
        \sum_{\bQ}v(\bQ+\bqbar)\,
        \lambda^{\mu\mu'}_{nm,n'm'}(\bkbar,\bkbar',\bqbar,\bQ)
        \right)
        \\
        \times\,
        d^\dagger_{s\mu n}(\bkbar+\bqbar)\,
        d^\dagger_{s'\mu'n'}(\bkbar'-\bqbar)\,
        d_{s'\mu'm'}(\bkbar')\,
        d_{s\mu m}(\bkbar),
    \end{aligned}
\end{equation}
where \(N_{\bkbar}\) is the number of sampled mBZ momenta and \(\bQ\) is a reciprocal lattice vector of the electron-crystal superlattice. The band operators are related to the plane-wave basis through
\begin{equation}
    d_{s\mu n}^\dagger(\bkbar)
    =
    \sum_{\alpha,\bG}
    C_{\mu\alpha s\bG,n}(\bkbar)\,
    c^\dagger_{\mu\alpha s}(\bkbar+\bG),
\end{equation}
with \(\alpha\) a combined layer-sublattice index and \(\bG\) a reciprocal vector of the superlattice. The projected form factors are
\begin{equation}
    \begin{aligned}
        \lambda^{\mu\mu'}_{nm,n'm'}(\bkbar,\bkbar',\bqbar,\bQ)
        =
        \sum_{\alpha\alpha'}\sum_{\bG\bG'}
        &C^*_{\mu\alpha s\,\bG+\bQ,n}(\bkbar+\bqbar)\,
        C^*_{\mu'\alpha' s'\,\bG'-\bQ,n'}(\bkbar'-\bqbar)
        C_{\mu'\alpha' s'\bG',m'}(\bkbar')\,
        C_{\mu\alpha s\bG,m}(\bkbar).
    \label{eq:form}
    \end{aligned}
\end{equation}
Because the interaction is spin independent, the spin labels are carried trivially by the external operators and the nontrivial momentum and orbital structure is encoded in \(\lambda\).

We solve the projected problem in the Green's-function language. Using the composite index \(\eta=(s,\mu,n)\), the single-particle Green's function is
\begin{equation}
    [G(\bkbar,\tau)]_{\eta\eta'}
    =
    -\langle
    T_\tau d_{\bkbar,\eta}(\tau)d^\dagger_{\bkbar,\eta'}(0)
    \rangle,
\end{equation}
and it satisfies the Dyson equation
\begin{equation}
    G^{-1}(\bkbar,i\omega_n)
    =
    i\omega_n+\mu-H_0(\bkbar)-\Sigma(\bkbar,i\omega_n),
\end{equation}
where \(H_0(\bkbar)\) is the projected single-particle Hamiltonian. In the Hartree--Fock approximation the self-energy is static, \(\Sigma(\bkbar,i\omega_n)\equiv\Sigma(\bkbar)\). The density matrix is
\begin{equation}
    \rho_{\eta\eta'}(\bkbar)
    =
    \langle
    d^\dagger_{\bkbar,\eta'}d_{\bkbar,\eta}
    \rangle,
\end{equation}
and the total self-energy is decomposed into Hartree and Fock contributions,
\(\Sigma=\Sigma^H+\Sigma^F\), with
\begin{equation}
    [\Sigma^H(\bkbar)]_{\eta\eta'}
    =
    \frac{1}{N_{\bkbar}}
    \sum_{\bkbar',\zeta,\zeta',\bQ}
    v(\bQ)\,
    \Lambda^H_{\eta\eta';\zeta\zeta'}(\bkbar,\bkbar';\bQ)\,
    \rho_{\zeta'\zeta}(\bkbar'),
\end{equation}
\begin{equation}
    [\Sigma^F(\bkbar)]_{\eta\eta'}
    =
    -\frac{1}{N_{\bkbar}}
    \sum_{\bkbar',\zeta,\zeta',\bQ}
    v(\bkbar'-\bkbar+\bQ)\,
    \Lambda^F_{\eta\zeta;\zeta'\eta'}(\bkbar,\bkbar';\bQ)\,
    \rho_{\zeta'\zeta}(\bkbar').
\end{equation}
The tensors \(\Lambda^{H/F}\) are obtained from the same projected form factors,
\begin{equation}
    \Lambda^H_{\eta\eta';\zeta\zeta'}(\bkbar,\bkbar';\bQ)
    =
    \lambda^{\mu\mu'}_{nm,n'm'}(\bkbar,\bkbar',\bqbar=0,\bQ),
\end{equation}
\begin{equation}
    \Lambda^F_{\eta\zeta;\zeta'\eta'}(\bkbar,\bkbar';\bQ)
    =
    \lambda^{\mu\mu'}_{nm,n'm'}(\bkbar,\bkbar',\bqbar=\bkbar'-\bkbar,\bQ),
\end{equation}
with the composite-index assignments
\(\eta=(s,\mu,n)\),
\(\eta'=(s,\mu,m)\),
\(\zeta=(s',\mu',m')\), and
\(\zeta'=(s',\mu',n')\). 

The self-consistent loop is implemented as follows. Starting from a trial density matrix with randomly chosen flavor occupations, we evaluate \(\Sigma^H\) and \(\Sigma^F\), construct the static Hartree--Fock Hamiltonian
\begin{equation}
    H_{\mathrm{HF}}(\bkbar)
    =
    H_0(\bkbar)+\Sigma^H(\bkbar)+\Sigma^F(\bkbar),
\end{equation}
diagonalize it to obtain an updated density matrix, and iterate with mixing until both the density matrix and the self-energy converge. The zero-temperature Hartree--Fock energy is then
\begin{equation}
    E_{\mathrm{HF}}
    =
    \sum_{\bkbar}
    \mathrm{Tr}
    \left[
    \left(
    H_0(\bkbar)+\frac{1}{2}\Sigma(\bkbar)
    \right)
    \rho(\bkbar)
    \right].
\end{equation}
To visualize the charge modulation, we reconstruct the real-space density from the converged Hartree--Fock Bloch states,
\begin{equation}
    n(\mathbf{r})
    =
    \sum_{\bG}n(\bG)e^{i\bG\cdot\mathbf{r}},
\end{equation}
with
\begin{equation}
    n(\bG)
    =
    \frac{1}{N_{\bkbar}}
    \sum_{\bkbar}\sum_{n m\in \mathrm{occ}}
    \langle
    u_n^{\mathrm{HF}}(\bkbar)
    |
    u_m^{\mathrm{HF}}(\bkbar+\bG)
    \rangle,
\end{equation}
where $\ket{u_n(\bkbar)^\text{HF}}$ is the periodic part of the $n$-th converged Hartree Fock wavefunction. In the calculations underlying this model description, we retain five conduction bands, keep four shells of plane waves, and freeze the remote valence bands. This truncation is appropriate for the relatively large displacement fields of interest, where the low-energy conduction bands are well separated from the valence sector. We provide representative HF bands, density profile and Berry curvature distributions for WC phase, AHC phase and hAHC phase mentioned in the main text in Fig.~\ref{figs2}.

\section{Time Dependent Hartree Fock theory and gRPA equations}

In this appendix we review the time dependent Hartree Fock theory (TDHF) and gRPA equations. We also guide the readers to Ref.~\cite{ambuj2025phonon,shao25elecmag} for more details. The collective-mode spectrum of the AHC shown in the main text is computed by solving gRPA equations which extends the static mean-field framework of the previous appendix to small fluctuations about the self-consistent ground state.
We start by examining the equation of motion of the density matrix,
\begin{align}
    i\hbar\partial_t \hat{\rho}(t) &= [\hat{H}_{\rm MF}[\hat{\rho}], \hat{\rho}(t)],\\
\hat{H}_{\text{MF}}[\hat{\rho}]  &= \hat{T} + \hat{\Sigma}^{H}[\hat{\rho}]+ \hat{\Sigma}^{F}[\hat{\rho}].
\end{align}
Consider a small perturbation on the density matrix
\begin{equation}
\hat{\rho}(t) = \hat{\rho}_{0} + \delta\hat{\rho}(t).
\end{equation}
After substituting this linear expansion into the equation of motion and preserving the leading order terms
\begin{equation}
    i\hbar \partial_t \delta\hat{\rho}(t) = [H_{\rm MF}[\hat{\rho}_0], \delta\hat{\rho}(t)] +[\delta \Sigma, \hat{\rho}_0],
\end{equation}
where we use the identity $[H_{\rm MF}[\rho_0],\rho_0]=0$. Furthermore, the above equation can be written under the eigen basis of the Hartree Fock mean field Hamiltonian,
\begin{equation}
    i\hbar \partial_t \delta\hat{\rho}_{mn}(t) = [H_{\rm MF}[\hat{\rho}_0], \delta\hat{\rho}(t)]_{mn} +[\delta \Sigma, \hat{\rho}_0]_{mn}.
    \label{eq:eom}
\end{equation}
For the AHC and WC cases we study, the one-body density matrix in the projected band basis is
\begin{equation}
\rho^{\mu\mu'}_{mn}(\bkbar,\bqbar,t)
=\left<d^\dagger_{n\mu'}(\bkbar)\,d_{m\mu}(\bkbar+\bqbar)\right>,
\end{equation}
and the first-order self-energy variations are
\begin{align}
(\delta\Sigma_{H})^{ss''}_{mn}(\bk,\bq)
&=
\delta_{ss''}\sum_{\bk',m',n',s'}
V^{ss'}_{mm'n'n}(\bk,\bk',\bq)\,
\rho^{s's'(1)}_{n'm'}(\bk'-\bq,\bq),\\
(\delta\Sigma_{F})^{ss'}_{mn'}(\bk,\bq)
&=
-\sum_{\bk',m',n}
V^{ss'}_{mm'n'n}(\bk',\bk,\bq+\bk-\bk')\,
\rho^{ss'(1)}_{nm'}(\bk'-\bq,\bq),
\end{align}
where
\begin{equation}
V^{ss'}_{mm'n'n}(\bk,\bk',\bq)
=
\frac{1}{N_{\bk}}\sum_{\bQ}
v(\bq+\bQ)\,
\lambda^{ss'}_{nm,n'm'}(\bk,\bk',\bq,\bQ).
\end{equation}
Note that $\mu$ and $s$ are both generalized indices that represents the spin or valley and the form factor $\lambda$ is defined as Eq.~\ref{eq:form}.
Then the detailed form of Eq.~\ref{eq:eom} should be
\begin{equation}
\begin{aligned}
i\hbar\,\frac{\partial}{\partial t}\rho^{\mu\mu'(1)}_{ph}(\bkbar,\bqbar)
={}&(\epsilon^{\mu}_{p\bkbar+\bqbar}-\epsilon^{\mu'}_{h\bkbar})\,\rho^{\mu\mu'(1)}_{ph}(\bkbar,\bqbar)\\
&+\sum_{\bkbar',p',h'}\Bigl[
\sum_{s'}\delta_{\mu\mu'}V^{\mu s'}_{p\bkbar+\bqbar\,h'\bkbar'\,p'\bkbar'+\bqbar\,h\bkbar}\rho^{s's'(1)}_{p'h'}(\bkbar',\bqbar)
-V^{\mu\mu'}_{p\bkbar+\bqbar\,h'\bkbar'\,h\bkbar\,p'\bkbar'+\bqbar}\rho^{\mu\mu'(1)}_{p'h'}(\bkbar',\bqbar)\Bigr]\\
&+\sum_{\bkbar',p',h'}\Bigl[
\sum_{s'}\delta_{\mu\mu'}V^{\mu s'}_{p\bkbar+\bqbar\,p'\bkbar'-\bqbar\,h'\bkbar'\,h\bkbar}\rho^{s's'(1)}_{h'p'}(\bkbar'-\bqbar,\bqbar)
-V^{\mu\mu'}_{p\bkbar+\bqbar\,p'\bkbar'-\bqbar\,h\bkbar\,h'\bkbar'}\rho^{\mu\mu'(1)}_{h'p'}(\bkbar'-\bqbar,\bqbar)\Bigr].
\end{aligned}
\end{equation}
where we use $\rho^{(1)}$ to represent the small fluctuation. The companion equation for $\rho^{(1)}_{hp}$ follows from the Hermiticity relation $\rho^{(1)}_{hp}=(\rho^{(1)}_{ph})^{*}$.

Inter-flavor particle--hole excitations, i.e. spin- and valley-reversal excitations, are decoupled from the intra-flavor charge excitations. Because the crystalline order develops only in the majority flavor, we now focus on the majority-flavor sector.

\begin{figure}
\includegraphics[width=0.9\columnwidth]{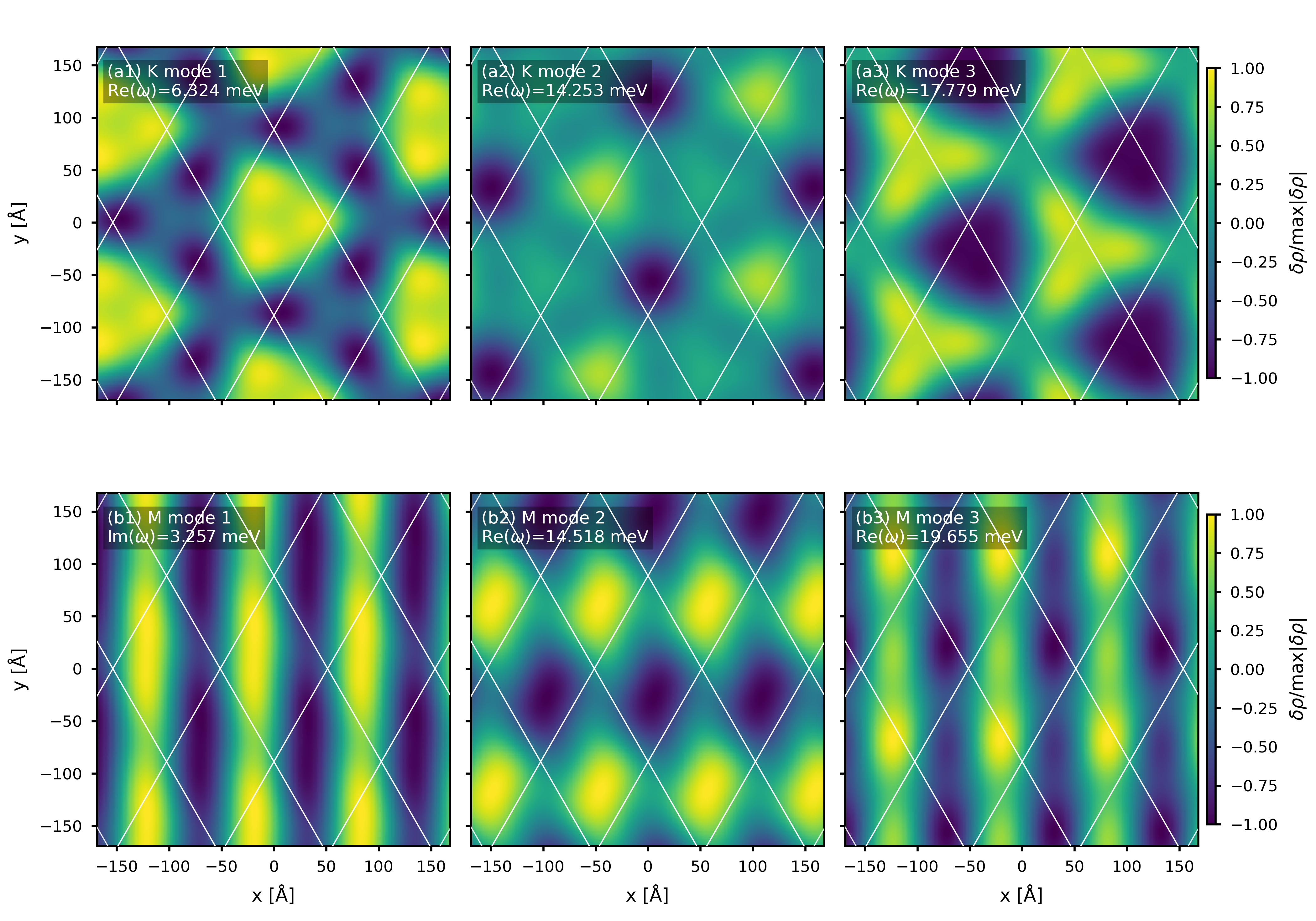}
\caption{
Density modulation of collective modes at high-symmetry momenta for the anomalous Hall crystal state at
$n=1.1\times10^{12}\,\mathrm{cm}^{-2}$ and $D=0.05\,\mathrm{eV}$.
Panels (a1)--(a3) show the three lowest selected modes at the $K$ point, while
panels (b1)--(b3) show the corresponding modes at the $M$ point.
The plotted quantity is the normalized real-space density fluctuation
$\delta\rho(\mathbf{r})/\max|\delta\rho|$, reconstructed from the gRPA eigenvectors
and shown over the superlattice spanned by $\mathbf{L}_1$ and $\mathbf{L}_2$.
Mode energies are indicated in each panel; imaginary-frequency modes are labeled by
$\mathrm{Im}(\omega)$, signaling a TDHF instability.
}
\label{figs3}
\end{figure}
The converged Hartree--Fock solution preserves the spin and valley quantum numbers of each band, so that $(s,\mu)$ is uniquely determined by the band index $p$ (or $h$) and the explicit upper indices can be suppressed. To track the residual spin-conservation constraints we introduce a generalized Kronecker symbol $\delta^{0}_{p'h'}$ which equals one when bands $p'$ and $h'$ carry the same flavor and zero otherwise. With this convention the dynamical matrices are
\begin{align}
A_{ph,p'h'}(\bkbar,\bkbar',\bqbar)
&= (\epsilon_{p\bkbar+\bqbar}-\epsilon_{h\bkbar})\,\delta_{pp'}\delta_{hh'}\delta_{\bkbar\bkbar'}
+ \delta^{0}_{ph}\delta^{0}_{p'h'}V^{s_{h}s_{h'}}_{p\bkbar+\bqbar\,h'\bkbar'\,p'\bkbar'+\bqbar\,h\bkbar}
- \delta^{0}_{hh'}\delta^{0}_{pp'}V^{s_{p}s_{h}}_{p\bkbar+\bqbar\,h'\bkbar'\,h\bkbar\,p'\bkbar'+\bqbar},\\
B_{ph,p'h'}(\bkbar,\bkbar',\bqbar)
&= \delta^{0}_{ph}\delta^{0}_{p'h'}V^{s_{h}s_{h'}}_{p\bkbar+\bqbar\,p'\bkbar'-\bqbar\,h'\bkbar'\,h\bkbar}
- \delta^{0}_{ph'}\delta^{0}_{p'h}V^{s_{p}s_{h}}_{p\bkbar+\bqbar\,p'\bkbar'-\bqbar\,h\bkbar\,h'\bkbar'},
\label{eq:ABmatrix}
\end{align}
and the linearized equations take the compact form
\begin{equation}
\begin{aligned}
i\hbar\,\frac{\partial}{\partial t}\rho_{ph}^{(1)}(\bkbar,\bqbar)
&= \sum_{\bkbar',p',h'}A_{ph,p'h'}(\bkbar,\bkbar',\bqbar)\,\rho_{p'h'}(\bkbar',\bqbar)
+ B_{ph,p'h'}(\bkbar,\bkbar',\bqbar)\,\rho_{h'p'}(\bkbar'-\bqbar,\bqbar),\\
-i\hbar\,\frac{\partial}{\partial t}\rho_{hp}^{(1)}(\bkbar-\bqbar,\bqbar)
&= \sum_{\bkbar',p',h'}A^{*}_{ph,p'h'}(\bkbar,\bkbar',-\bqbar)\,\rho_{h'p'}(\bkbar'-\bqbar,\bqbar)
+ B^{*}_{ph,p'h'}(\bkbar,\bkbar',-\bqbar)\,\rho_{p'h'}(\bkbar',\bqbar),
\end{aligned}
\end{equation}
where we used the Hermiticity relation $\rho^{ss'(1)*}_{ph}(\bkbar,\bqbar)=\rho^{s's}_{hp}(\bkbar+\bqbar,-\bqbar)$ and relabelled $\bqbar\to-\bqbar$. We now seek oscillating solutions
\begin{align}
\rho_{ph}^{(1)}(\bkbar,\bqbar,t)
&= X_{ph}(\bkbar,\bqbar)\,e^{-i\omega t} + Y^{*}_{ph}(\bkbar,\bqbar)\,e^{i\omega^{*}t},\\
\rho_{hp}^{(1)}(\bkbar+\bqbar,-\bqbar,t)
&= Y_{ph}(\bkbar,\bqbar)\,e^{-i\omega t} + X^{*}_{ph}(\bkbar,\bqbar)\,e^{i\omega^{*}t},
\end{align}
which retain both positive- and negative-frequency components and so manifestly satisfy $\rho^{(1)}_{hp}=(\rho^{(1)}_{ph})^{*}$. Inserting this ansatz and matching coefficients of $e^{-i\omega t}$ yields
\begin{align}
\hbar\omega\,X_{ph}(\bkbar,\bqbar)
&= \sum_{\bkbar',p',h'}A_{ph,p'h'}(\bkbar,\bkbar',\bqbar)\,X_{p'h'}(\bkbar',\bqbar)
+ B_{ph,p'h'}(\bkbar,\bkbar',\bqbar)\,Y_{p'h'}(\bkbar',-\bqbar),\\
-\hbar\omega\,Y_{ph}(\bkbar,-\bqbar)
&= \sum_{\bkbar',p',h'}A^{*}_{ph,p'h'}(\bkbar,\bkbar',-\bqbar)\,Y_{p'h'}(\bkbar',-\bqbar)
+ B^{*}_{ph,p'h'}(\bkbar,\bkbar',-\bqbar)\,X_{p'h'}(\bkbar',\bqbar),
\end{align}
which can be cast in the compact matrix form
\begin{equation}
\hbar\omega\begin{pmatrix} X \\ Y \end{pmatrix}
=
\begin{pmatrix} A & B \\ -B^{*} & -A^{*} \end{pmatrix}
\begin{pmatrix} X \\ Y \end{pmatrix}
\equiv R \begin{pmatrix} X \\ Y \end{pmatrix}
\label{eq:gRPA}
\end{equation}
This is the generalized RPA (gRPA) eigenvalue problem stated in the main text. The structure of the kernels in Eq.~\eqref{eq:ABmatrix} immediately implies that spin-conserving and spin-flip channels decouple: the spin-projection factors $\delta^{0}_{ph}$ and $\delta^{0}_{p'h'}$ restrict each block to be either same-flavor or opposite-flavor particle--hole pairs.

For a gRPA eigenmode $\nu$ at wave vector $\vec q$, the associated charge-density fluctuation is obtained from
\begin{equation}
\delta n_{\nu \vec{q}}(\vec r)
=
2\,\Re\!\Bigl[
\sum_{p,h,\bkbar} X_{ph}^\nu(\bkbar,\bqbar)\,\psi_{p\bkbar+\bqbar}(\mathbf{r})\,\psi^{*}_{h\bkbar}(\mathbf{r})
+ Y_{ph}^\nu(\bkbar,-\bqbar)\,\psi_{h\bkbar}(\mathbf{r})\,\psi^{*}_{p\bkbar-\bqbar}(\mathbf{r})
\Bigr],
\end{equation}
where $\psi_{n\bkbar}(\mathbf{r})$ is the Hartree--Fock Bloch wavefunction. A typical density modulation of AHC at high symmetry momenta is shown in Fig.~\ref{figs3}. We clearly see the stripe features induced by the soft mode at the $M$ point.

Imaginary eigenfrequencies of $R$ signal a dynamical instability of the underlying Hartree--Fock state, while the corresponding eigenvector identifies the unstable charge density pattern through $\delta n_{\nu \vec{q}}(\vec r)$. This is the criterion used in the main text to map out the dynamically unstable region of the AHC and to identify the soft transverse phonon mode responsible for bulk--edge coupling.

\section{Effective edge theory after integrating out bulk transverse phonons}

The purpose of this Appendix is to show that integrating out the low-lying bulk phonons in AHC produces a Luttinger-liquid-like effective theory for the edge modes. 
We consider a strip geometry:
the system occupies the region $0<y<W$, with edges parallel to $\hat{x}$. The bottom edge is located at $y_b=0$, and the top edge is located at $y_t=W$. The bottom edge is taken to be right-moving, while the top edge is left-moving. 

Let $\phi_b(x,t)$ and $\phi_t(x,t)$ be the chiral boson fields on the bottom and top edges and their actions are given by,
\begin{equation}
S_{\rm edge}
=
-\sum_{\eta=b,t}
\frac{1}{4\pi}
\int dt\,dx\,
\left[
s_\eta
\left(\partial_x\phi_\eta\right)
\left(\partial_t\phi_\eta\right)
+
v_\eta
\left(\partial_x\phi_\eta\right)^2
\right].
\end{equation}
Here the chirality index $
s_b=+1$ and $ s_t=-1$.
With this convention, the equation of motion gives the chiral dispersion
$\omega_\eta(q_x)=s_\eta v_\eta q_x$. Thus the bottom edge has positive-energy modes for $q_x>0$, $
\omega_b(q_x)=v_b q_x$, whereas the top edge has positive-energy modes for $q_x<0$,
$\omega_t(q_x)=-v_t q_x $.
We now describe the long-wavelength phonons using elastic theory by expanding the free-energy in terms of the displacement vector,
\begin{equation}
\mathbf{u}(\mathbf{r},t)
=
\left(
u_x(\mathbf{r},t),
u_y(\mathbf{r},t)
\right),
\end{equation}
The Fourier mode of the displacement vector is
\begin{equation}
u_i(\mathbf{r},t)
=
\int \frac{d\omega dq_x dq_y}{(2\pi)^3} 
e^{i(q_xx+q_yy-\omega t)}
u_i(q_x,q_y,\omega),
\end{equation}
and the Gaussian action for the bulk phonons is
\begin{equation}
S_{\rm ph}
=
\frac{1}{2}
\int \frac{d\omega dq_x dq_y}{(2\pi)^3} 
u_i(-q_x,-q_y,-\omega)
\left[
D^{-1}(q_x,q_y,\omega)
\right]_{ij}
u_j(q_x,q_y,\omega). 
\end{equation}
Here \(D_{ij}\) is the displacement Green's function.
As shown in the main text, the edge mode mainly intersects with the transverse phonon continuum.  In elastic theory, a purely transverse deformation has no compressional strain, $\nabla\cdot \mathbf u_T=0$, so its leading coupling to the edge density is naturally through shear strain 
$u_{xy}=\frac{1}{2}
\left(\partial_x u_y+\partial_y u_x\right)$. Then, the bulk-edge coupling is taken to be
\begin{equation}
S_{\rm int}
= g \sum_{\eta=b,t}
\int dt\,dx\,
\rho_\eta(x,t)
u_{xy}(x,y_\eta,t),
\end{equation}
where  $y_b=0$ and $y_t=W$. Substituting the Fourier expansion gives
\begin{equation}
S_{\rm int}
=
\sum_{\eta=b,t} \int
\frac{d\omega dq_x dq_y}{(2\pi)^3} 
\rho_\eta(-q_x,-\omega)
g_\eta
C_i(q_x,q_y)
e^{iq_y y_\eta}
u_i(q_x,q_y,\omega).
\end{equation}
The phase factor \(e^{iq_y y_\eta}\) comes from evaluating the bulk phonon field at the edge position \(y=y_\eta\)
and we used Fourier transformation of shear-strain, $u_{xy}(q_x,q_y,\omega)
=
C_i(q_x,q_y)
u_i(q_x,q_y,\omega),$
with the shear-strain vertex $C_i(q_x,q_y)
=
\frac{i}{2}(q_y,q_x)$. Next, we integrate out the phonons by writing $S_{\rm int}$ as
\begin{equation}
S_{\rm int}
=
\int \frac{d\omega dq_x dq_y}{(2\pi)^3} 
J_i(-q_x,-q_y,-\omega)
u_i(q_x,q_y,\omega).
\end{equation}
\begin{equation}
J_i(-q_x,-q_y,-\omega)
=
\sum_{\eta=b,t}
\rho_\eta(-q_x,-\omega)
g_\eta
C_i(q_x,q_y)
e^{iq_y y_\eta}.
\end{equation}
Integrating out the displacement field $u_i$ leads to the effective interaction,
\begin{align}
S_{\rm ind}
&=
-\frac{1}{2}
\int \frac{d\omega dq_x dq_y}{(2\pi)^3}
J_i(-q_x,-q_y,-\omega)
D_{ij}(q_x,q_y,\omega)
J_j(q_x,q_y,\omega) \\
&=
-\frac{1}{2}
\sum_{\eta,\eta'=b,t}
\int \frac{d\omega dq_x}{(2\pi)^2}
\rho_\eta(-q_x,-\omega)
U_{\eta\eta'}(q_x,\omega)
\rho_{\eta'}(q_x,\omega),
\end{align}
where the phonon-mediated edge state interaction is
\begin{equation}
U_{\eta\eta'}(q_x,\omega)
=
g^2
\int \frac{dq_y}{2\pi}
C_i(q_x,q_y)
D_{ij}(q_x,q_y,\omega)
C_j(q_x,q_y)^*
e^{iq_y(y_\eta-y_{\eta'})}.
\end{equation}

The resulting effective edge theory is therefore
\begin{equation}
    S_{\rm eff}=S_{\rm edge}+ S_{\rm ind} + S_{\rm Coulomb}
\end{equation}
where $S_{\rm Coulomb}$ is the usual instantaneous Coulomb interaction between edge densities. Since the direct Coulomb interaction is expected to dominate the intra-edge density-density interaction, we can neglect the intra-edge part of $S_{\rm ind}$. The resulting theory is therefore a Luttinger-liquid variant: left- and right-moving edge densities are coupled as in an ordinary Luttinger liquid, but the coupling between opposite edges is retarded.

The phonon-mediated inter-edge interaction is,
\begin{equation}
U_{bt}^R(q_x,\omega)
=
g^2
\int \frac{dq_y}{2\pi}
C_i(q_x,q_y)
D_{ij}^{R}(q_x,q_y,\omega)
C_j(q_x,q_y)^*
e^{-iq_y W}.
\end{equation}
The factor $e^{-iq_y W}$ records the transverse propagation from one edge to the other. Thus the large-$W$ behavior is controlled by the analytic structure of the retarded phonon propagator in the complex $q_y$ plane. If the integrand is smooth on the real $q_y$ axis, the oscillatory phase suppresses the integral at large $W$. However, if the edge excitation is on shell with a propagating bulk mode, the $q_y$ integral can receive a pole contribution that is not exponentially suppressed by $W$.
To see this more clearly, suppose the retarded displacement Green's function has the following form,
\begin{equation}
D_{ij}^{R}(q_x,q_y,\omega)=
\frac{Z_{ij}(q_x,q_y)}
{(\omega+i0^+)^2-\omega_T^2(q_x,q_y)},
\end{equation}
where $Z_{ij}$ contains the polarization and spectral weight of the transverse mode. Suppose the on-shell condition $\omega
=\omega_T(q_x,k_y)$ has a real solution $k_y$, we then expand near the pole,
\begin{equation}
\omega_T(q_x,q_y)
\simeq
\omega
+
v_y^T(q_x,k_y)
(q_y-k_y),
\end{equation}
\begin{equation}
v_y^T(q_x,k_y)
=
\left.
\frac{\partial \omega_T(q_x,q_y)}
{\partial q_y}
\right|_{q_y=k_y}.
\end{equation}
Computing the integral using residue theorem, we find
\begin{equation}
U_{bt}^R(q_x,\omega)
=
-i g^2
\sum_{k_y}
\frac{
C_i(q_x,k_y)
Z_{ij}(q_x,k_y)
C_j(q_x,k_y)^*
}
{
2\omega
\left|
v_y^T(q_x,k_y)
\right|
}
e^{-ik_y W}.
\end{equation}
The sum is over the retarded outgoing-wave solutions for propagation between the two edges. Since $k_y$ is real, $|e^{-ik_yW}|=1$. Thus the interaction may oscillate with $W$, but it is not exponentially suppressed by the strip width. Its amplitude is instead controlled by the phonon spectral weight, the edge-phonon vertex, and the transverse group velocity. If the on-shell condition has no real transverse momentum solution, the most general solution then has the form $k_y=q_y'-i\kappa$, with $\kappa>0$, and the inter-edge interaction contains the evanescent factor $e^{-\kappa W}$. Therefore the distinction between a propagating and an evanescent inter-edge interaction is determined by whether the edge excitation lies inside the bulk transverse-phonon continuum. 

Let us now consider an ideal isotropic acoustic transverse phonon where the propagator is given by the familiar form,
\begin{equation}
D_{ij}^{T,R}(q,\omega)
=
\frac{\delta_{ij} -\frac{q_iq_j}{q^2}}
{\rho_M\left[
(\omega+i\delta)^2-c_T^2(q_x^2+q_y^2)
\right]},
\end{equation}
Here $c_T$ is the transverse sound velocity and $\rho_M$ is the spring constant. In this case, the inter-edge coupling becomes
\begin{equation}
U_{bt}^R(q_x,\omega)
=
\frac{g^2}{\rho_M}
\int \frac{dq_y}{2\pi}
\frac{
\frac{1}{4}
\frac{(q_x^2-q_y^2)^2}
{q_x^2+q_y^2}
}
{
(\omega+i0^+)^2
-
c_T^2(q_x^2+q_y^2)
}
e^{-iq_y W}.
\end{equation}
The on-shell condition is simply
$\omega^2
=c_T^2(q_x^2+q_y^2)$. For $\omega<c_Tq_x$, the transverse momentum is purely imaginary, $q_y=\pm i\kappa$,  then the two edges are decoupled in the thermodynamic limit since  $U_{bt}^R(q_x,\omega) \propto e^{-\kappa W}$. However, when $\omega>c_Tq_x$, the transverse momentum is real,
\begin{equation}
q_y=\pm k_y,
\qquad
k_y=
\sqrt{
\frac{\omega^2}{c_T^2}
-
q_x^2
}.
\end{equation}
The retarded pole then gives an outgoing propagating wave. For propagation from the top edge to the bottom edge, we have
\begin{equation}
U_{bt}^R(q_x,\omega)
=
-i
\frac{g^2}
{2\rho_M c_T^2 k_y}
\left[
\frac{1}{4}
\frac{(q_x^2-k_y^2)^2}
{q_x^2+k_y^2}
\right]
e^{i k_y W}.
\end{equation}

At zero wave-vector $q_x=0$, the retarded inter-edge interaction takes a  simple form,
\begin{equation}
U_{bt}^R(q_x=0,\omega)
=
\frac{-ig^2 k_y}
{8\rho_M c_T^2}
e^{i k_y W}=
\frac{-ig^2}
{8\rho_M c_T^3} \,\omega\,
e^{i \omega W/c_T}.
\end{equation}
Here $W/c_T$ is the time required for the transverse mode to propagate across the sample width. Thus, the inter-edge coupling is not exponentially suppressed by $W$; instead it carries the propagation phase accumulated by a transverse wave traveling between the two edges. We also note that $U_{bt}^R(q_x=0,\omega)$ vanishes linearly with $\omega$ and this limit commutes with first taking $\omega=0$ then $q_x\rightarrow 0$.

\begin{figure}
\includegraphics[width=1\columnwidth]{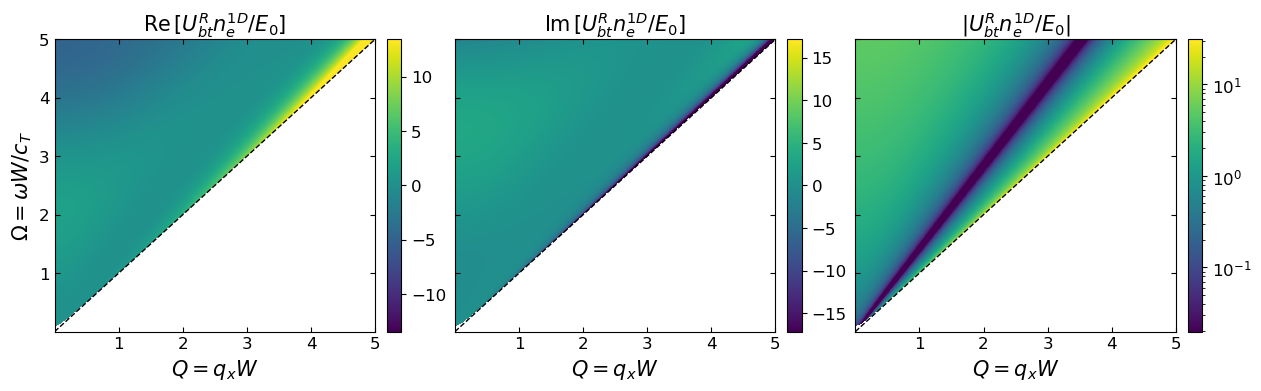}
\caption{
Inter-edge coupling mediated by an isotropic acoustic transverse phonon.
The dashed line marks $\Omega=Q$, the threshold above which the transverse momentum $K=\sqrt{\Omega^2-Q^2}$ is real and the bulk mode propagates between opposite edges. 
In this propagating regime, the inter-edge interaction is not exponentially suppressed by the strip width, but instead carries the phase $e^{iK}$ accumulated across the sample. 
Note $U_{bt}=0$ along $\Omega=\sqrt{2}Q$ arises from the simple shear-strain form factor.
}
\label{fig_sumpmat}
\end{figure}

To conclude our discussion, let us plot the inter-edge coupling in the $q_x$--$\omega$ plane. We first define the energy scale
\begin{equation}
E_0
=
\frac{g^2 n_e^{1D}}
{8\rho_M c_T^2 W},
\end{equation}
where $n_e^{1D}$ is the one-dimensional electron density along the edge. In terms of the dimensionless variables $Q=q_xW$ and $\Omega=\frac{\omega W}{c_T}$,
the inter-edge coupling can be written as
\begin{equation}
\frac{U_{bt}^R(Q,\Omega)n_e^{1D}}{E_0}
=
-i
\frac{
e^{i\sqrt{\Omega^2-Q^2}}
}
{
\sqrt{\Omega^2-Q^2}
}
\frac{
(2Q^2-\Omega^2)^2
}
{\Omega^2},
\end{equation}
valid in the propagating regime $\Omega>Q$. Here $Q$ measures the edge wavelength relative to the sample width. For an edge density wave with wavelength $\lambda_x=2\pi/q_x$, one has $Q=2\pi W/\lambda_x$. The dimensionless frequency $\Omega$ measures the frequency in units of the transverse phonon crossing time $W/c_T$. For a realistic micron-scale device, the strip width is typically $W\sim 1$--$10\,\mu{\rm m}$. If a microwave structure excites edge density waves with wavelength $\lambda_x\sim 5$--$50\,\mu{\rm m}$, then a natural range is $Q\sim 0.1$--$10$. Similarly,
\begin{equation}
\Omega
=
2\pi f\frac{W}{c_T}
\simeq
3.1
\left(
\frac{f}{1\,{\rm GHz}}
\right)
\left(
\frac{W}{5\,\mu{\rm m}}
\right)
\left(
\frac{10^4\,{\rm m/s}}{c_T}
\right).
\end{equation}
Thus GHz frequencies naturally correspond to $\Omega\sim 1$--$10$ for micron-scale devices with $c_T\sim 10^4\,{\rm m/s}$. In Fig.~\ref{fig_sumpmat}, we show the dimensionless inter-edge coupling in the $Q$--$\Omega$ plane. The line $\Omega=Q$ marks the threshold of the propagating transverse-phonon continuum. Above this line, the transverse momentum is real and the coupling carries the propagation phase across the sample. The coupling also vanishes along $\Omega=\sqrt{2}Q$ in this simple isotropic model, which follows from the shear-strain vertex. This zero is a form-factor effect rather than a universal feature.
 
\section{Estimation of the bulk edge coupling constant}

In this subsection, we provide a simple estimate of the bulk-edge coupling constant $g$ in this equation
\begin{equation}
S_{\rm int} = g
\sum_{\eta={\rm B},{\rm T}}
\int dt\,dx\,
\rho_{\eta}(x,t)\,
u_{xy}(x,y_{\eta},t).
\end{equation}
Following Ref.~\cite{giustino2017electron}, the electron-phonon coupling  $g$ can be understood as the quasiparticle energy shift produced by a strain tensor $u_{xy}$,
\begin{equation}
g_\eta(k_x)\sim
\frac{\partial \epsilon_{\eta k_x}}{\partial u_{xy}} .
\end{equation}
where
\begin{equation}
\delta \epsilon_{\eta k_x}
=
\left\langle
\chi_{\eta k_x}
\left|
\delta\Phi_{\rm strain}
\right|
\chi_{\eta k_x}
\right\rangle .
\end{equation}
and $\delta\Phi_{\rm strain}$ is the change of the self-consistent crystal potential to the linear order in the $u_{xy}$ and $\chi_{\eta k_x}$ is the wavefunction of the edge states.

For simplicity, we assume the self-consistent Hartree-Fock potential can be written as a local periodic form,
\begin{equation}
\Phi(\mathbf r)
=
\sum_{\mathbf G}
\Phi_{\mathbf G}
e^{i\mathbf G\cdot \mathbf r},
\label{eq:HF_potential_shear}
\end{equation}
 A small displacement field shifts the electronic-crystal texture according to
\begin{equation}
\Phi(\mathbf r)
\rightarrow
\Phi(\mathbf r-\mathbf u).
\end{equation}
To linear order in $\mathbf u$, the induced change of the potential is
\begin{equation}
\delta \Phi(\mathbf r)
=
-\mathbf u(\mathbf r)\cdot\nabla\Phi(\mathbf r)=
-i
\sum_{\mathbf G}
\left[
\mathbf G\cdot\mathbf u(\mathbf r )
\right]
\Phi_{\mathbf G}
e^{i\mathbf G\cdot\mathbf r}.
\label{eq:delta_phi_displacement}
\end{equation}

We now consider a displacement profile $u(\br) = u_{T\bq}\mathbf{e}_T(\bq) e^{i\bq\cdot \br}$, which is induced by a long-wavelength ($q\ll \sqrt{n}$, $n$ is the doping density) transverse phonon mode. More specifically, we consider a transverse bulk phonon propagating perpendicular to the edge,
\begin{equation}
\mathbf q=q_y\hat{\mathbf y},
\qquad
\mathbf e_T=\hat{\mathbf x}.
\end{equation}
Then we have 

\begin{equation}
\delta \Phi(\mathbf r) =
-i
\sum_{\mathbf G}
\left( G_x u_{T q_y} e^{iq_y y}
\right)
\Phi_{\mathbf G}
e^{i\mathbf G\cdot\mathbf r}.
\end{equation}
The corresponding edge-state energy shift is
\begin{equation}
\delta \epsilon_{\eta k_x}
=
\left\langle
\chi_{\eta k_x}
\left|
\delta\Phi_{\rm strain}
\right|
\chi_{\eta k_x}
\right\rangle .
\end{equation}
Therefore,
\begin{equation}
\delta \epsilon_{\eta k_x}
=-i u_{T q_y}  \sum_\bG G_x \Phi_\bG \left\langle
\chi_{\eta k_x}
\left|
e^{iq_y y+i\bG \cdot \br}
\right|
\chi_{\eta k_x}
\right\rangle
\end{equation}
The coupling constant can then be estimated,
\begin{align}
g
=&
-2i
\left.
\frac{\partial^2 \delta \epsilon_{\eta k_x}}
{\partial u_{T q_y}\,\partial q_y}
\right|_{q_y=0}\\
 =& -2i\sum_\bG G_x \Phi_\bG \left< \chi_{\eta k_x} \left| y e^{i\bG \cdot \br} \right| \chi_{\eta k_x}\right>.
\end{align}
The matrix elements can be estimated as,
\begin{equation}
\left\langle
\chi_{\eta k_x}
\left| y
e^{i\mathbf G\cdot\mathbf r}
\right|
\chi_{\eta k_x}
\right\rangle
\sim
\xi_\eta
\mathcal F^\eta_{\mathbf G},
\end{equation}
where $\mathcal F^\eta_{\mathbf G}$ is a dimensionless edge form factor and $\xi_\eta$ is in the length scale and characterize the width of the edge state wavepacket. With this, the coupling constant is given by
\begin{equation}
|g|\sim
\xi_\eta
\left|
\sum_{\mathbf G}
G_x \Phi_{\mathbf{G}}\mathcal{F}^\eta_{\mathbf G}
\right|\sim
G_1\xi_\eta |\Phi_{G_1}\mathcal{F}^\eta_{G_1}|.
\end{equation}
where we have only kept the first shell of reciprocal lattice above.
The effective potential $\Phi_{G_1}$ can be estimated from the Hartree-Fock Bragg gap $\Delta_{\rm Bragg}$ opened by the electronic-crystal potential. Including a dimensionless edge-bulk overlap factor $\mathcal O_\eta<1$, $|\Phi_{G_1}\mathcal{F}^\eta_{G_1}|
\sim
\Delta_{\rm Bragg}\mathcal O_\eta.$
Therefore,
\begin{equation}
|g|
\sim
G_1\xi_\eta
\Delta_{\rm Bragg}
\mathcal O_\eta.
\label{eq:g_estimate_final}
\end{equation}
For the anomalous Hall crystal studied in the main text, the ordered phase appears near $n\sim 10^{12}\ {\rm cm}^{-2}$. Taking $\xi_\eta\sim 1\ {\rm nm}$, $\Delta_{\rm Bragg}\sim 10\ {\rm meV}$, and $G_1\xi_\eta=\mathcal O(1)$, the shear coupling is naturally estimated to be $|g|\sim 1\text{--}10\ {\rm meV}$, depending on the edge-bulk overlap factor $\mathcal O_\eta$.

\end{document}